\documentclass[onecolumn,showpacs,preprintnumbers,amsmath,amssymb]{revtex4}
\usepackage{graphicx,psfrag,hyperref}
\usepackage{dcolumn}
\usepackage{bm,bbm}
\usepackage{xcolor}
\begin{document}

\title{Aligning a Majorana fermion's anapole moment with an external current through photon emission mediated by the fermion's generalized polarizabilities}

\author{Kiana Walter}
\author{Kobi Hall}
\author{David C. Latimer}

\affiliation{Department of Physics, University of Puget Sound,
Tacoma, WA 98416-1031
}

\newcommand*{\sech}{\mathop{\mathrm{sech}}\limits}
\newcommand*{\balpha}{\boldsymbol{\alpha}} 
\newcommand*{\dilog}{\mathrm{Li}_2}
\newcommand{\qslash}{\not{\hbox{\kern-3pt $q$}}}
\newcommand{\kslash}{\not{\hbox{\kern-3pt $k$}}}
\newcommand{\pslash}{\not{\hbox{\kern-2pt $p$}}}
\newcommand{\delslash}{\not{\hbox{\kern-3pt $\partial$}}}
\newcommand{\Dslash}{\not{\hbox{\kern-3pt $D$}}}
\newcommand{\gmn}{g^{\mu \nu}}
\newcommand{\Pslash}{\not{\hbox{\kern-2.3pt $P$}}}
\newcommand{\Kslash}{\not{\hbox{\kern-2.3pt $K$}}}
\newcommand{\Pslashsup}{^\not{\hbox{\kern-0.5pt $^P$}}}
\newcommand{\Poddup}{^\not{\hbox{\kern-0.5pt $^\mathcal{P}$}}}
\newcommand{\be}{\begin{equation}}
\newcommand{\ee}{\end{equation}}
\newcommand{\al}[1]{\begin{align}#1\end{align}}
\newcommand{\gv}{\gamma^5}
\newcommand{\bsig}{\boldsymbol{\sigma}}
\def\rcurs{{\mbox{$\resizebox{.16in}{.08in}{\includegraphics{ScriptR}}$}}}
\def\brcurs{{\mbox{$\resizebox{.16in}{.08in}{\includegraphics{BoldR}}$}}}
\def\hrcurs{{\mbox{$\hat \brcurs$}}}
\newcommand{\brp}{\mathbf{r}_+}
\newcommand{\brm}{\mathbf{r}_-}
\newcommand{\diff}{\mathrm{d}}
\newcommand{\bs}{\bigskip}
\newcommand{\beps}{\boldsymbol{\epsilon}}

\begin{abstract}
The sole static electromagnetic property of a spin-$\frac{1}{2}$ Majorana fermion is its anapole moment. Though they cannot couple to single real photons, these particles can interact with electric currents through virtual photons. If a Majorana fermion is immersed in a background current, there is an energy difference between the spin states of the fermion; the higher energy state has its anapole moment antialigned with the current. 
  In this paper, we address the ability of a system of initially unpolarized Majorana fermions to achieve some degree of polarization relative to a static background current.  In considering processes that allow the Majorana fermion's spin to flip to the lower-energy state, we focus upon two irreversible processes:  the spontaneous emission of two real photons and the emission of a single real photon emitted in virtual Compton scattering.  Both of these processes involve coupling to photons via the fermion's polarizaibilities.  We compute the spin-flip transition rates for these processes using a low-energy expansion of the Hamiltonian and construct a toy model to showcase how these rates depend upon the underlying parameters within a model.  Applying these ideas to a thermal dark matter (DM) model, we find that when the DM thermally decouples from the Standard Model plasma in the early universe, two-photon emission is negligible but partial polarization for the DM medium can proceed via virtual Compton scattering if sufficient currents exist.  
  
  \end{abstract}

\maketitle

\section{Introduction}

The self-conjugate nature of Majorana fermions severely restricts their coupling to the electromagnetic field.  For a spin-$\frac{1}{2}$ Majorana fermion, its charge, electric dipole moment, and magnetic dipole moment must vanish identically. Its sole static electromagnetic property is the anapole moment \cite{bk82, nieves}.  A classical model of the anapole exists: a toroidal current distribution; see, for example, Ref.~\cite{gray_karl_novikov}.  This system has a nonzero anapole moment along its symmetry axis, and additionally, its charge and electric and magnetic dipole moments vanish.  As a result, the classical torus does not interact with static electric or magnetic fields, 
but it does experience a torque in the presence of an external current density, which acts to align the torus's anapole moment with the current.  This notion carries over faithfully to the realm of particle physics.  Majorana fermions do not couple to single real photons for the same reasons that the classical torus does not respond to electric or magnetic fields.  But, as with the torus, the interaction Hamiltonian between a fermion's anapole moment, $\mathbf{a}$, aligned with the particle's spin and an external current density, $\mathbf{J}$, is given by $H_\text{anapole} = - \mathbf{a} \cdot \mathbf{J}$ \cite{zeldovich}.   
From this Hamiltonian, we see that particles with anapole moments aligned with a background external current are in a lower energy state than those with moments antialigned with the current.
The background current results in an energy difference between these two states by an amount $\Delta E = 2 a J$. 

Suppose a system of initially unpolarized identical spin-$\frac{1}{2}$ Majorana fermions were  subjected to a uniform constant current density.  Assuming sufficient interaction, the system will come into thermal equilibrium after some time and exhibit a degree of polarization according to the Boltzmann distribution, $n(\uparrow)/n(\downarrow) = \exp(\Delta E/T)$ (assuming spin-up fermions are aligned with the current).  The time scale over which this occurs is dependent upon the interaction rate for processes that admit a spin flip of the Majorana fermion.  The dominant mode of interaction is through the exchange of a single photon with the current, which, at the individual particle level,  is a reversible process.  The ability to polarize the system via such reversible processes crucially relies upon sufficient interactions with the environment to achieve thermalization. But, in addition to reversible interactions, there are also effectively irreversible processes that allow the Majorana fermion to flip to the lower-energy state with  the anapole moment aligned with the external current. We will study these subdominant irreversible processes in this paper and take up the  reversible anapole interactions in subsequent work.  

For a Majorana fermion with anapole moment antialigned with an external current, photon emission is one irreversible mechanism in which the fermion can flip its spin to the lower energy state; however, spontaneous emission of a {\em single} photon is not possible because anapoles cannot couple to  a  single real photon.  
But a Majarona fermion can couple to two real photons via its model-dependent electric and magnetic polarizabilities \cite{radescu}.  Through these polarizabilities, spontaneous emission of two photons is possible.  

Generically, a spin-$\frac{1}{2}$ fermion has sixteen independent polarizabilties, which can be classified by how the two-photon interaction transforms under parity ($\mathcal{P}$) and time reversal ($\mathcal{T}$) transformations.  There are six polarizabilities that are separately invariant under $\mathcal{P}$- and $\mathcal{T}$-transformations \cite{prange} and an additional four polarizabilities that are $\mathcal{P}$-odd and $\mathcal{T}$-even \cite{chen}; the remaining six polarizabilities are $\mathcal{P}$-odd and $\mathcal{T}$-odd \cite{gorchtein}.  The Majorana fermion's self-conjugate nature constrains this list somewhat, requiring the four $\mathcal{P}$-odd and $\mathcal{T}$-even polarizabilities to vanish \cite{maj_2photon}. This leaves a dozen modes through which Majorana fermions can interact with two real photons. 
In this work, we will focus upon only those which are $\mathcal{P}$- and $\mathcal{T}$-even because the other couplings are likely to be suppressed (given that they require $\mathcal{P}$- and $\mathcal{T}$- violation).  Of these six $\mathcal{P}$- and $\mathcal{T}$-even polarizabilities, two are spin independent -- the familiar electric and magnetic polarizabilites, while four are spin dependent \cite{ragusa,ragusa2}.  It is these spin-dependent polarizabilities, that will admit a spin flip of the Majorana fermion via the spontaneous emission of two real photons.

The study of particle polarizabilities has a long history in nuclear physics in the context of real Compton scattering (RCS) \cite{prange, jacob_mathews, hearn, bardeen, Bernard:1991rq, hhkk, babusci, holstein_hopol, bedaque, chen, Bernard:2002pw, Hildebrandt:2003md, holstein_sumrules, gorchtein} because photons can be used as a clean probe of nucleon structure.  More recently, virtual Compton scattering (VCS) has emerged as an additional probe \cite{guichon,vanderhaeghen,scherer,metz,pasquini,drechsel,drechsel2,HHKS,HHKD,lvov,gorchtein_vcs}.  In VCS, the incoming photon that interacts with the target nucleon is virtual with four momentum satisfying $q^2 = - Q^2 <0$ rather than $q^2=0$ as in RCS.  In the low energy limit, one may describe VCS via an effective interaction using generalized polarizabilities which are now functions of the initial photon's four momentum, $Q^2$.  Focusing upon only the $\mathcal{P}$- and $\mathcal{T}$-even interactions, the ostensibly ten generalized polarizabilities \cite{guichon} can be reduced to six after applying charge conjugation and crossing symmetries \cite{drechsel2}.  These six terms should match up with the RCS polarizabilities in the limit in which the initial photon becomes real, but care must be taken in how this limit is realized.  The low energy expansion of the VCS amplitude developed by Gorchtein \cite{gorchtein_vcs} provides a scheme that continuously connects the VCS and RCS polarizabilities. 

Given the presence of a background current, a Majorana fermion can undergo single photon emission by coupling to charged particles through its generalized polarizabilities in VCS.  The charged particles can interact with the Majorana fermion via a virtual photon, and the Majorana fermion can emit a real final-state photon in the process.  As with RCS, some of the generalized polarizabilites are spin dependent and, thus, able to effect a spin flip.  We will compute the transition rate for this process as well as  spontaneous two-photon emission.

In what follows, we first introduce the effective anapole interaction between the Majorana fermion and background current and construct a toy model to show how this interaction emerges from a more complete theory.  
We then consider the effective interaction between two real photons and the Majorana fermion.  Focusing upon RCS in the low-energy limit, we compute the Majorana fermion's polarizabilities. 
With these, we determine the spontaneous two-photon emission rate for a fermion undergoing a spin flip to the lower energy state.
Finally we move on to a discussion of the low-energy expansion of the VCS amplitude.  We follow Gorchtein's scheme in Ref.~\cite{gorchtein_vcs} to determine the generalized polarizabilities for our toy model.  Gorchtein uses the Breit frame, so we shift the results over to the lab frame in order to allow us to compute the interaction rate for spin-flip under single photon emission.  This process is not spontaneous, so we also determine the rate at which the Majorana fermion's anapole moment transitions from aligned to antialigned with the current through  VCS. 

The ability of a current to polarize a collection of Majorana fermions is intrinsically interesting, but it could have consequential relevance to neutrino physics or dark matter (DM) models \cite{anapole_dm1,anapole_dm2,Anchordoqui,chua,kumar,ahmed,neves2021majorana}.  Because we only consider Majorana fermions that are non-relativistic, our results are not directly applicable to neutrinos, though generalizing our results to relativistic particles could be an interesting avenue to pursue.   But, for DM thermal relics, once the universe cools to the point at which the DM decouples from the plasma of Standard Model particles, local electric currents could result in regions of polarized DM that would persist to present day.  Using reasonable parameters for the toy model, we estimate the rate at which two-photon emission and VCS occur around the time that DM decouples from the thermal bath.  We find that two-photon emission is essentially forbidden in this era, but the VCS process could be appreciable in the presence of a sufficiently large, yet still non-relativistic, current.

This possibility has implications for indirect DM detection experiments. In such experiments, the detection of high-energy Standard Model particles 
could be ascribed to DM annihilation, if the detected particles have no other astrophysical origin.  Majorana fermions can rather generically annihilate into photons if they are in an $s$-wave state  \cite{anapole_2photon,chua}, though sometimes this mode is naturally suppressed  \cite{kumar,anapole_dm1,anapole_dm2}.   Regardless, non-observation of high-energy photons from DM annihilation results in an upper bound on the annihilation cross section.  However, if a region of DM were partially polarized in the early universe by a background current, this polarization would persist to the present, and the polarization would suppress $s$-wave annihilation because this mode requires the  two annihilating DM particles to have opposite spins.  In such a region, an indirect detection experiment would result in an overly stringent bound on the annihilation cross section if the analysis assumed an unpolarized DM medium.

\section{Anapole moment and toy model \label{toy}}

\begin{figure}[h]
\includegraphics[height=4cm]{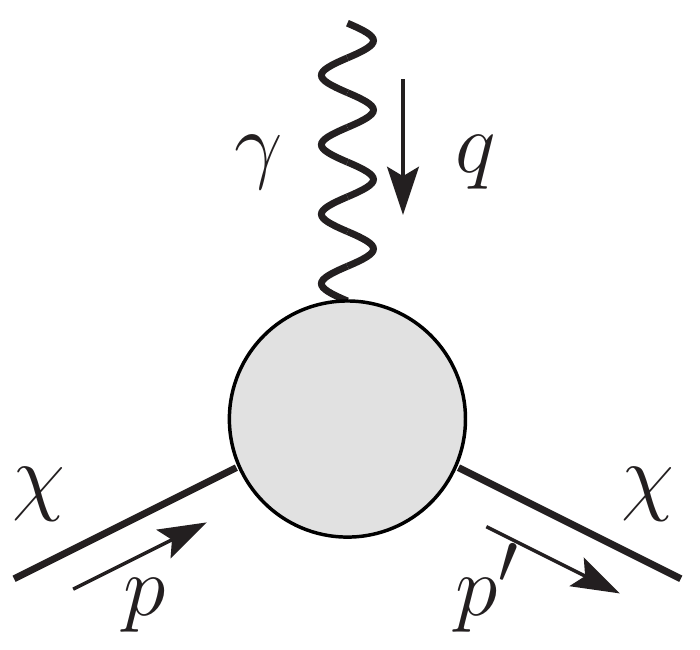}
\caption{ Anapole vertex for a Majorana fermion.  \label{fig1}}
\end{figure}

When considering the electromagnetic interactions of a Majorana fermion, $\chi$, we take an effective field theory approach, expressing the Lagrangian density in terms of the electromagnetic field strength tensor, $F^{\mu \nu} = \partial^\mu A^\nu - \partial^\nu A^\mu$. For the single-photon interaction represented in Fig.~\ref{fig1}, the effective Lagrangian is 
\begin{equation}
\mathcal{L}_\text{anapole} = f_\text{a} (q^2) \bar{\chi} \gamma^\mu \gamma^5 \chi \partial^\nu F_{\mu \nu}.  \label{L_ana}
\end{equation}
The static anapole moment is determined in the limit of vanishing photon momentum, $q\to 0$; that is, we set $f_\text{a}:=f_\text{a}(q^2=0)$. A few remarks are in order.  First, reflecting the fact that Eq.~\ref{L_ana} is an effective Lagrangian, the anapole moment has mass dimension $[M]^{-2}$.
Second, it is clear from the structure of the Lagrangian term that the interaction vanishes for real, transverse photons given that $q^2=0$ and $\epsilon \cdot q$, where $\epsilon^\mu$ is the photon's polarization vector, viz. $A^\mu = \epsilon^\mu e^{-i q \cdot x}$.  Finally, the derivative of the field tensor can be replaced with an external density $J^\mu = \partial_\nu F^{\mu\nu}$, highlighting the coupling between the Majorana fermion's anapole moment and an external current.

For our calculations, we will be working in the low-energy limit.  The Majorana fermion will be non-relativistic, $|\mathbf{p}|, |\mathbf{p}'|\ll m_\chi$, and momentum transfer will be small $|\mathbf{q}| \ll m_\chi$. In this regime, the full Dirac four-spinor is not needed to describe the particle dynamics.  In the low energy limit, the spatial parts of the Lagrangian, Eq.~(\ref{L_ana}), dominate resulting in the anapole interaction Hamiltonian    
\begin{equation}
H_\text{anapole} = -f_a \boldsymbol{\sigma} \cdot \mathbf{J},
\end{equation}
which acts on Pauli spinors, $\xi$.  In this two state system, the energy difference between spinors antialigned and aligned with the current is $\omega_0 = 2 f_a J$.
 
A particle's anapole moment $f_a$ is a model-dependent quantity whose parameter dependences can only be determined from a more fundamental theory.   It is useful to consider a toy model to see how model parameters enter into the effective coupling for the anapole moment and for the polarizabilities that follow.   For our toy model, we suppose the Majorana fermion couples to a charged scalar, $\phi$, and a Dirac fermion, $\psi$.  In order for the anapole moment to be nonzero, the coupling must be parity violating. For our toy model, we assume maximal parity violation.  Explicitly, we take the interaction term in the Lagrangian density to be
\begin{equation}
\mathcal{L}_\text{int} = g \bar{\psi}\frac{1}{2}(1 - \gamma^5) \chi  \phi^*+ \mathrm{h.c.}   \label{L_int}.
\end{equation}
To simplify our expressions, we will also assume that the mass of the scalar particle dominates the fermion masses.

\begin{figure}[h]
\includegraphics[height=4cm]{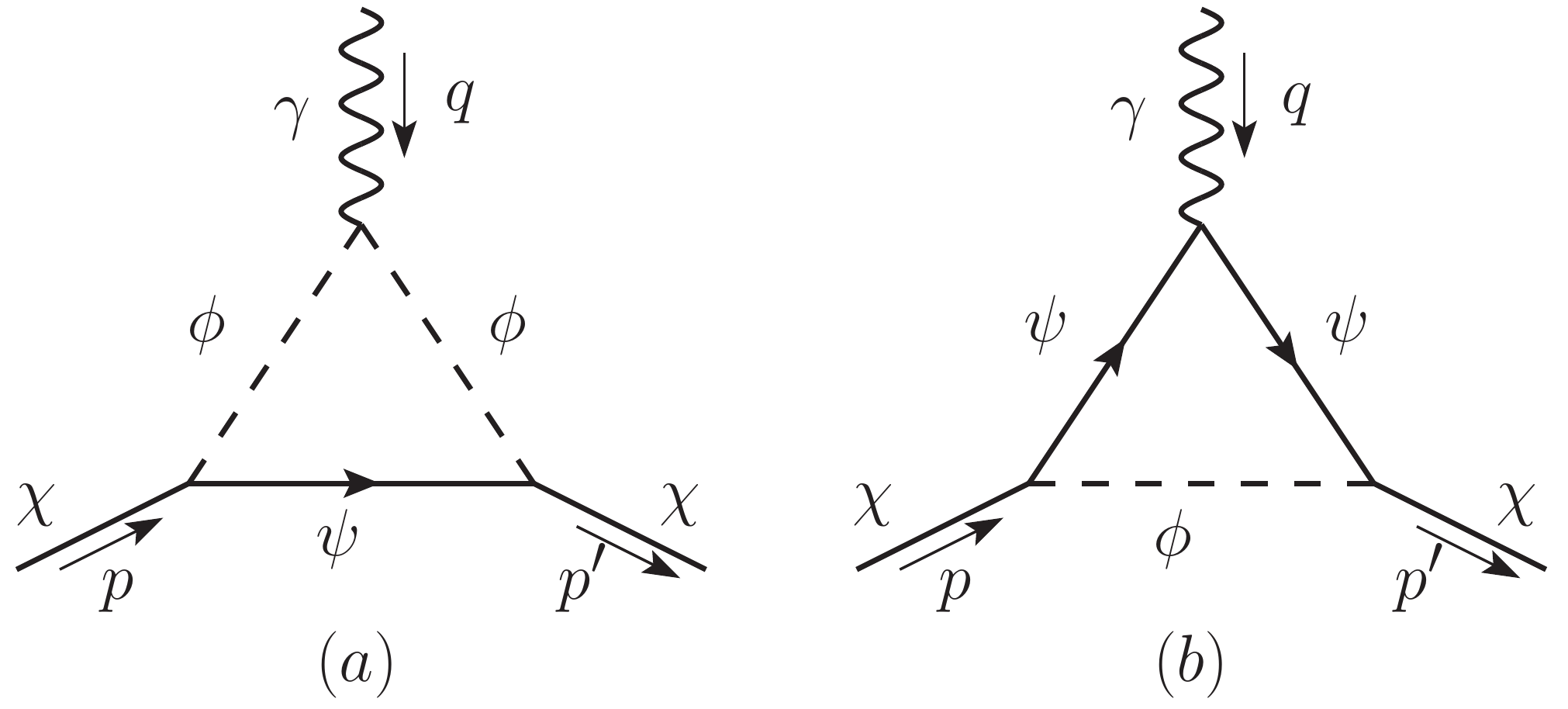}
\caption{  One-loop diagrams that determine the Majorana fermion's anapole moment.  \label{fig2}}
\end{figure}

The anapole moment can be determined from the one-loop Feynman diagrams in Fig.~\ref{fig2}.  For the computation, we use the Feynman rules for Majorana fermions developed in Refs.~\cite{denner1,denner2}.  In particular, we must consider both orientations of the Dirac fermion flow in loop diagrams.   
Per the effective Lagrangian in Eq.~(\ref{L_ana}), the anapole vertex has the form $i f_a (q^2 \gamma^\mu - \qslash q^\mu) \gamma^5$; therefore, to determine the anapole moment, it is sufficient to compute the diagrams in Fig.~\ref{fig2} out to order $\mathcal{O}(q^2)$. We execute the calculations in {\tt Mathematica} using {\tt Package-X 2.0} developed in Ref.~\cite{patel}.  The resulting moment is then given by
\begin{equation}
f_a \approx \frac{eg^2}{(4\pi)^2M_\phi^2} \left[ \frac{1}{3} \log\left( \frac{M_\phi^2}{m_\psi^2}  \right)- \frac{1}{2} \right], \label{fa}
\end{equation}
where we keep only the leading terms in $M_\phi \gg m_\chi, m_\psi$.   We note that the structure of this anapole moment is similar to the charge radius of the neutrino in the analogous limits \cite{neutrino_anapole}.

\section{Spontaneous two-photon emission}

\subsection{Effective Hamiltonian}
\begin{figure}[h]
\includegraphics[height=4cm]{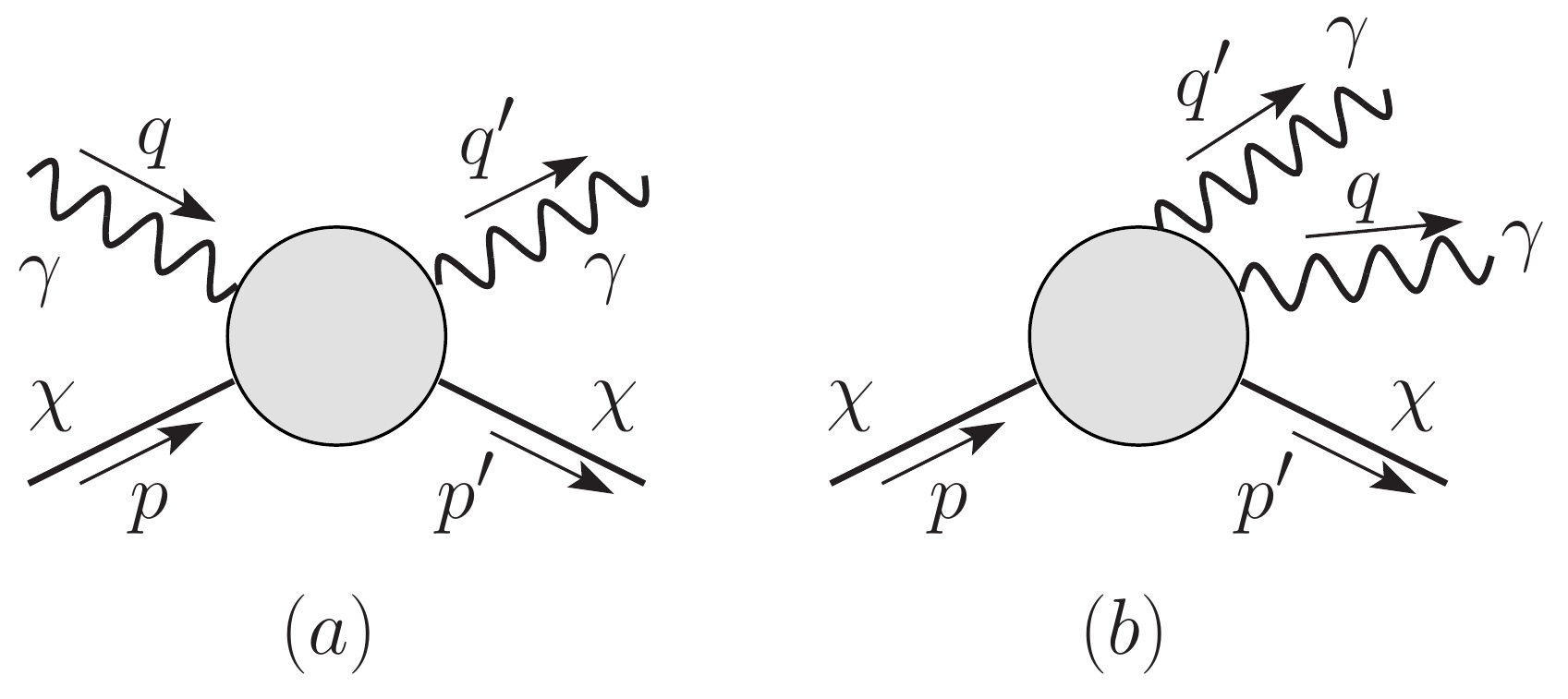}
\caption{ Two-photon interactions with a Majorana fermion by virtue of the particle's polarizability. (a) Real Compton scattering. (b) Two-photon emission.  \label{fig3}}
\end{figure}

To compute the interaction of the Majorana fermion with real photons, we introduce an effective Hamiltonian that couples two photons via the fermion's polarizabilities.
 At low energies, the interactions can be expressed rather simply in terms of the photons' electric and magnetic field components.
For the $\mathcal{P}$- and $\mathcal{T}$-invariant interactions, two of these interaction terms are spin independent, while the remaining four are spin dependent and involve derivatives of the fields.  Following the parametrization for the spin-dependent terms laid out in Ragusa \cite{ragusa,ragusa2}, we write the low-energy expansion (LEX) of the Hamiltonian as 
\al{
H_\text{pol} =&    -\frac{1}{2} 4\pi(\alpha_E \mathbf{E} \cdot \mathbf{E} + \beta_M \mathbf{B} \cdot \mathbf{B}) \nonumber \\
& + 4\pi  \tfrac{1}{2} \gamma_1 \mathbf{E} \cdot[ \bsig \times (\boldsymbol{\nabla} \times \mathbf{B}) ] - 4\pi  (\gamma_2 +\gamma_4) \mathbf{B} \cdot [\boldsymbol{\nabla}(\bsig \cdot \mathbf{E})] \nonumber \\
&  + 4\pi  \gamma_3 \mathbf{E} \cdot [\boldsymbol{\nabla}(\bsig \cdot \mathbf{B})]   +4\pi \left( \tfrac{1}{2} \gamma_2 + \gamma_4\ \right) \mathbf{B} \cdot[ \bsig \times (\boldsymbol{\nabla} \times \mathbf{E}) ], \label{H_pol}
}
where the coefficients represent the various polarizabilities.
This Hamiltonian is appropriate both for real Compton scattering, as in Fig.~\ref{fig3}(a), and  the emission of two photons, Fig.~\ref{fig3}(b).

Particle polarizabilities can be easily understood within the context of classical electromagnetism.  In a two-photon interaction with the Majorana fermion, one can heuristically think of one photon's electric or magnetic field inducing an electric or magnetic dipole with which the second photon can interact.  The simplest examples are the spin-independent polarizabilities. For instance, in the presence of an electric field, an electromagnetic object will typically deform; the positively charged constituents feel a tug in the direction of the field while the negatively charged bits feel a tug in the opposite direction. To leading order, the field induces an electric dipole in the same direction of the field, $\mathbf{p} =  \alpha_E \mathbf{E}$. This dipole will lower the energy of the system via its interaction with the field $\sim - \mathbf{p} \cdot \mathbf{E}$, resulting in a potential term in the Hamiltonian $-\frac{1}{2} \alpha_E E^2$.  The contribution from the spin-independent magnetic polarizability, $\beta_M$, is similar, contributing to the system energy an amount $-\frac{1}{2} \beta_M B^2$.   Higher order interactions between the particle and (derivatives of) the electromagnetic field can result in spin-dependent polarizabilities \cite{ragusa,ragusa2}.  Because these moments involve derivatives of the fields, their contribution to the Hamiltonian will be $\mathcal{O}(\omega^3)$, with $\omega$ the photon energy.   For a spin-$\frac{1}{2}$ particle, the two leading order spin-dependent electric dipole moments are $-\frac{1}{2} \gamma_1\bsig \times (\boldsymbol{\nabla} \times \mathbf{B})$ and $ - \gamma_3 \boldsymbol{\nabla}(\bsig \cdot \mathbf{B})$, and the corresponding spin-dependent magnetic dipole moments are $ (\gamma_2 +\gamma_4) \bsig \times (\boldsymbol{\nabla}\times \mathbf{E})$ and $ -(\tfrac{1}{2} \gamma_2 +\gamma_4)\boldsymbol{\nabla} (\bsig\cdot \mathbf{E})$. The coefficients $\gamma_j$ are known as the spin-dependent polarizabilities. We note that our seemingly peculiar expressions for these coefficients are consistent with a prevailing  definition of the spin-dependent polarizabilities in the literature, e.g., in Ref.~\cite{hhkk}.

Before moving on to discuss two-photon emission, we will make a few comments on the RCS process.  Our reasons are twofold. First,  we will explicitly show that our expression is consistent with the leading order LEX of the RCS amplitude found in the literature.  Second, we would like to have an expression for the RCS Hamiltonian matrix element that can be more easily compared to that for the VCS Hamiltonian.
One challenge in the literature has been to show that the VCS result limits to the RCS case whenever the incident photon becomes real.  In Ref.~\cite{gorchtein_vcs}, Gorchtein finds an appropriate linear combination of fields such that the matrix element for the RCS Hamiltonian   is recovered from his expression for the VCS Hamiltonian in the appropriate limits.  We will show this for the non-Born terms in our work.  To facilitate this comparison, we will write a matrix element for the RCS Hamiltonian in terms of the photon momentum and polarization three-vectors, enforcing the transverse condition $\mathbf{q} \cdot \beps =0 = \mathbf{q}' \cdot \beps'$ and the on-shell condition $q^2 = 0 = q'^2$; that is, we will write the Hamiltonian in terms of the Fourier modes of the fields.  Our expression will be valid in both the center-of-mass frame and the Breit frame.  The Breit frame requires the incoming and outgoing fermion three-momenta satisfy $\mathbf{p}' = - \mathbf{p}$, and this fact forces the incident and outgoing photon energies to be the same $q^0= q'^0 =: \omega$.  

To set our notation, we work with the quantized vector field
\begin{equation}
A^\mu (x) = \sum_\lambda \int \frac{\diff^3 k}{(2\pi)^3} \frac{1}{\sqrt{2\omega}}\left[\epsilon^\mu(k,\lambda) a_{\mathbf{k},\lambda} e^{-i k \cdot x} + \epsilon^{*\mu} (k,\lambda) a^\dagger_{\mathbf{k},\lambda} e^{i k \cdot x}   \right]  , \label{vfield}
\end{equation}
where we sum over polarization states $\lambda$. 
We work in a gauge with $A^0 = 0$ so that the polarization vectors are purely spatial.
The electric and magnetic fields are found through the usual relations: $\mathbf{E} = -\frac{\partial}{\partial t} \mathbf{A}$ and $\mathbf{B} = \boldsymbol{\nabla} \times \mathbf{A}$.   For the RCS process, the initial and final real photons have momentum and polarization $(q, \epsilon)$ and $(q', \epsilon')$, respectively.  Omitting some normalization factors, we  express the RCS Hamiltonian matrix elements in terms of the electromagnetic field Fourier components
\al{
\langle q', \beps'| H_\text{pol} | q,\beps\rangle\sim & [ \omega^2 \alpha_E + (  \mathbf{q} \cdot \mathbf{q}'  )\beta_M   ]\beps \cdot \beps'^*    -\beta_M ( { \mathbf{q}} \cdot {\beps}'^*)  ( \mathbf{q}' \cdot \beps) 
\nonumber \\
&+i \omega  [   \omega ^2   \gamma_1 -   (\mathbf{q} \cdot \mathbf{q}')  (\gamma_2 + 2 \gamma_4  )] [ \bsig \cdot ({ \beps}'^{*} \times \beps)]   
\nonumber   \\
  &+i \omega  \gamma_2   (\beps'^* \cdot \beps) [\bsig \cdot( \mathbf{q}' \times \mathbf{q})]  \nonumber \\
& + i \omega   \gamma_4 \{       (\mathbf{q}' \cdot \beps ) [\bsig\cdot(\beps'^* \times \mathbf{q})]   -(\mathbf{q} \cdot \beps'^*) [\bsig\cdot(\beps \times \mathbf{q}')]    \} \nonumber \\
& +   i \omega \gamma_3  \{ ({\beps}   \cdot   \mathbf{q}' )[\bsig\cdot (\beps'^*\times \mathbf{q}' )] - (\beps'^{*}    \cdot    \mathbf{q}) [\bsig \cdot  (\beps \times   \mathbf{q})]  \}  . \label{rcs_ham}
}
This expression is consistent with the low-energy behavior of the RCS matrix element in Ref.~\cite{hhkk}, as applied to a Majorana fermion with vanishing charge and static dipole moments.

\subsection{Polarizabilities in the toy model}
\begin{figure}[h]
\includegraphics[height=4cm]{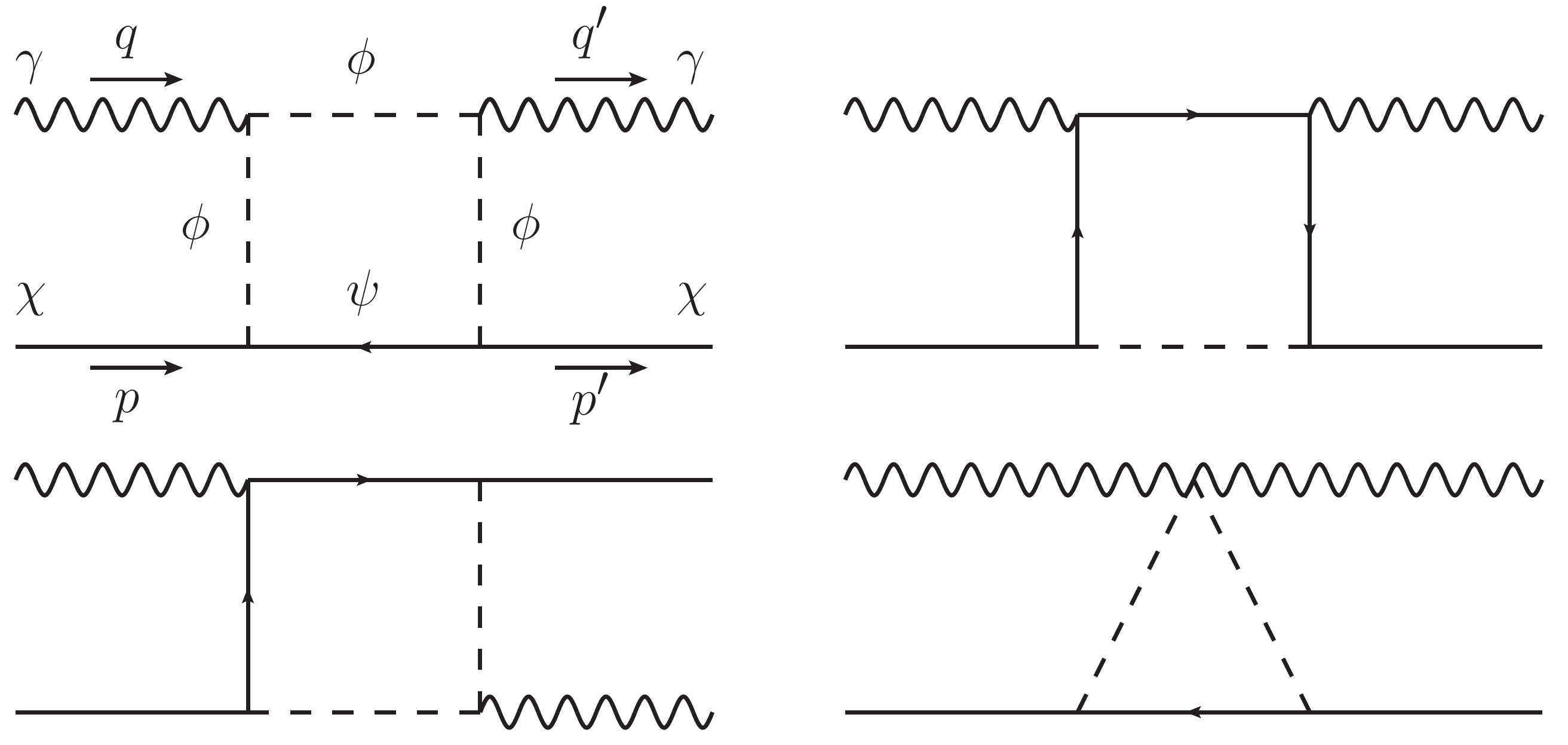}
\caption{ One-loop diagrams that contribute to the real Compton scattering amplitude.  Both orientations of fermion flow within the loop are considered, and the diagrams with the ``crossed" photons lines are not shown.  \label{fig4}}
\end{figure}

Polarizabilities are model-dependent quantities, so we turn again to the model introduced in Sec.~\ref{toy} to examine how model parameters can enter into these coefficients.  
A sample of the  one-loop Feynman diagrams that contribute to the RCS process is shown in Fig.~\ref{fig4}.  Again, we execute the calculations in {\tt Mathematica} using the package developed in Ref.~\cite{patel}.  We compute the amplitude to $\mathcal{O}(\omega^3)$ in the Breit frame and then extract the polarizability coefficients through comparison with the expression in Eq.~(\ref{rcs_ham}).
  Keeping only the leading order terms in this limit, we find the Majorana fermion's spin-independent polarizabilities, $\alpha_E$ and $\beta_M$, and spin-dependent polarizabilities, $\gamma_j$,  in the Breit (or CM) frame to be
\al{
\alpha_E \approx&  \frac{e^2g^2}{(4\pi)^3}  \frac{m_\chi}{M_\phi^4} \Bigg[ \frac{2}{3} \log\left(\frac{M_\phi^2}{m_\psi^2}\right)-  \frac{5}{6} \Bigg] \label{alpha}, \\
\beta_M \approx&  \frac{e^2 g^2}{(4\pi)^3} \frac{m_\chi}{M_\phi^4} \Bigg[\frac{2}{3} \log\left(\frac{M_\phi^2}{m_\psi^2}\right)    - \frac{13}{6}\Bigg], \\
 \gamma_1 \approx& - \frac{e^2g^2}{(4\pi)^3}  \frac{1}{3} 
 \frac{1}{M_\phi^2 m_\psi^2} \label{g1}, \\
\gamma_2 \approx&    \frac{e^2g^2}{(4\pi)^3}  \frac{1}{M_\phi^4} \Bigg[     \frac{2}{3} \log\left(\frac{M_\phi^2}{m_\psi^2}\right) -\frac{3}{2} \Bigg] \label{g2}, \\
\gamma_3 \approx& \frac{e^2g^2}{(4\pi)^3}
        \frac{1}{6}\frac{1}{M_\phi^2m_\psi^2} \label{g3},\\
\gamma_4 \approx&  - \frac{e^2g^2}{(4\pi)^3}   \frac{1}{6} \frac{1}{M_\phi^2 m_\psi^2}, \label{g4}  
}
where we consider the approximation in which the scalar mass dominates $m_\psi, m_\chi \ll M_\phi$. These values are consistent with the independent calculation in Ref.~\cite{maj_2photon}.

\subsection{Rate of spontaneous two-photon emission}

We now turn to the spontaneous emission of two photons, Fig.~\ref{fig3}(b), from a Majorana fermion immersed in a steady background current $\mathbf{J}$. We take the the current to be in the $z$-direction. Spin-down Majorana fermions, $\xi_-$, will have an energy of $\omega_0 = 2f_aJ$ greater than a spin-up, $\xi_+$, counterparts. We will assume that the Majorana fermion is at rest relative to the current both before {\em and} after photon emission; that is, we neglect recoil effects which are small because $m_\chi \gg \omega_0$.    
Shifting the calculation to the particle rest frame (from the previously discussed Breit or CM frame) has one additional implication. The polarizability coefficients that appear in the   low-energy Hamiltonian, Eq.~(\ref{H_pol}), for two-photon interactions are frame dependent; however, the corrections to the polarizabilities incurred  in boosting from the the Breit frame to the particle's rest frame are sub-dominant and thus negligible when considering only leading order terms.  Given this, we employ the polarizabilities calculated in Eqs.~(\ref{alpha}-\ref{g4}).

To compute the transition rate for spontaneous emission of two photons, we  use Fermi's golden rule
\begin{equation}
\diff^6 \Gamma_\text{spont} = 2\pi | \langle f | H_\text{pol}| i \rangle |^2  \rho_\text{f}.
\end{equation}
Here the initial state is a spin down Majorana fermion with $\mathbf{p}=0$, and the final state consists of a spin-up fermion with $\mathbf{p}'=0$ along with two photons with momenta and polarizations $q_1, q_2$ and $\beps_1^{(\alpha)}, \beps_2^{(\beta)} $, respectively.  For our calculations, we will sum over final photon polarizations, $\alpha$ and $\beta$, and 
integrate over the density of photon final states 
\begin{equation}
\rho_f = \prod_{j=1}^2 \frac{\diff^3 q_j}{(2\pi)^3}.
\end{equation}

Because the transition from initial to final state involves a spin flip for the Majorana fermion, the spin-independent terms in the Hamiltonian, Eq.~(\ref{H_pol}), have vanishing matrix element.  As such, there are four  contributions to the amplitude coming from the spin-dependent terms .
As an example, let us consider in some detail just the first spin-dependent electric polarizability: $H_{ \gamma_1} = 4\pi  \tfrac{1}{2} \gamma_1  \mathbf{E}\cdot[  \bsig \times (\boldsymbol{\nabla} \times \mathbf{B}) ]$.   The matrix element for this process is
\al{
\langle \text{f} | H_{{\gamma}_1} | \text{i} \rangle =& i \gamma_1 \frac{\pi}{\sqrt{\omega_1 \omega_2}} ( \omega_1 \omega_2^2-\omega_2 \omega_1^2) \bsig_\text{fi} \cdot \left({\beps_2^{(\beta)*}} \times {\beps_1^{(\alpha)*}}\right) ,
}
where we set $\bsig_\text{fi} = \langle \xi_+|\bsig | \xi_-\rangle$.   To compute the total transition rate for this, we integrate over the final phase space of the photons and sum over photon polarizations.  For a transition mediated solely by the $ \gamma_1$ term in Eq.~(\ref{H_pol}), we compute a spin-flip rate 
\begin{equation}
{\Gamma_\text{spont}}_{\gamma_1} = \frac{2 \gamma_1^2 {\omega_0}^9}{2835 \pi}.  
\end{equation}

The other contributions to two-photon spontaneous emission are computed in a similar fashion.
Generally, we must sum over these so that possible interfering terms can be given proper account.  After doing so (and summing over photon polarization states), we find 
the spontaneous two-photon emission transition rate to be
\begin{equation}
    \Gamma_\text{spont} =(2 \gamma_1^2 + 10 \gamma_2^2 - 3 \gamma_2 \gamma_3 + 12 \gamma_3^2 + 20 \gamma_2 \gamma_4 - 2 \gamma_3 \gamma_4 + 
   12 \gamma_4^2 + \gamma_1 \gamma_2 + 4  \gamma_1\gamma_3 + 3  \gamma_1\gamma_4) \frac{\omega_0^9}{2835 \pi}.
 \end{equation}

From the model-depend polarizabilities, Eq.~(\ref{g4}), we see that at leading order $ \gamma_2$ is smaller than the remaining spin-dependent coefficients by a factor of $m_\psi^2/M_\phi^2$.  For what remains, we define $\gamma :=  \gamma_3$ and find $ \gamma_1 = -2 \gamma$ along with $\gamma_4 = -\gamma$.  Neglecting $\gamma_2$, we can approximate the transition rate for this model as
\begin{equation}
   \Gamma_\text{spont} =  \frac{   32 \gamma^2 \omega_0^9}{2835  \pi}.  \label{spont_rate}
   \end{equation}

\section{Single-photon emission via VCS}

\subsection{Effective Hamiltonian}

\begin{figure}[h]
\includegraphics[height=4cm]{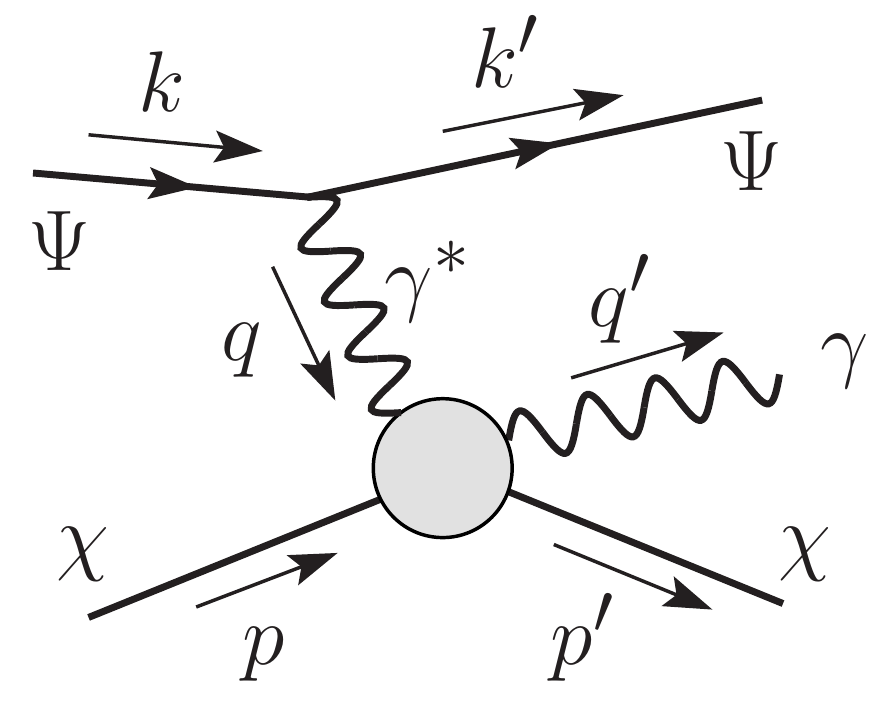}
\caption{ Single-photon emission from a Majorana fermion via virtual Compton scattering.  \label{fig5}}
\end{figure}

We now consider single-photon emission mediated by the interaction with a charged particle, coupled to the Majorana fermion via  generalized polarizabilties.  Though spontaneous emission of a single photon is not possible,
 a Majorana fermion can emit a single real photon in the virtual Compton scattering process depicted in Fig.~\ref{fig5}.  Here, the Majorana fermion interacts with a charged fermion, represented by $\Psi$ with charge $e$.  The amplitude for this process is 
 \begin{equation}
  \mathcal{M}_\text{VCS} = i e \bar{u}(k') \gamma_\mu u(k) \frac{1}{q^2} \mathcal{T}^{\mu\nu}_\text{VCS} \epsilon'^*_\nu(q',\lambda'), 
 \label{VCS_amp}
 \end{equation}
 where $\mathcal{T}^{\mu\nu}_\text{VCS}$ is the VCS tensor that encodes the physics within the shaded vertex in Fig.~\ref{fig5} and $q = k-k'$.  

 We can actually cast this amplitude in a more suggestive form by projecting the  Dirac bilinear from the charged fermion, $\Psi$,  onto a set of three orthogonal normalized vectors, $\epsilon^\mu(q,\lambda)$,
  \begin{equation}
\bar{u}(k') \gamma^\mu u(k) = \sum_\lambda \Omega_\lambda \epsilon^\mu(q,\lambda). \label{decomp}
 \end{equation}
To do so, we first note that the off-shell photon momentum, $q$, is orthogonal to Dirac bilinear, $\bar{u}(k') \qslash u(k) = 0$, because on-shell spinors satisfy $\kslash u(k) =m_\Psi u(k)$. 
 Given this, two of the of the vectors, $\epsilon^\mu(q,\lambda=\pm)$,  can be chosen to be wholly spatial.  Setting $q^\mu = (\omega, \mathbf{q})$, these two basis vectors are $\epsilon^\mu(q,\lambda=\pm) = (0, \beps_\pm)$ with the spatial part  orthogonal to the photon's three-momentum,  $\beps_\pm \cdot \mathbf{q}=0$.   For normalized $\beps_\pm$, then these two four-vectors have norm $\epsilon^\mu(q,\lambda=\pm) \epsilon_\mu(q,\lambda=\pm) =-1$. 
The third vector, $\epsilon^\mu(q,\lambda=0)$, must be time-like because  the off-shell photon's momentum is spatial, $q^2 := -Q^2 <0$, viz. $  |\mathbf{q}| > \omega$. 
 Then, the longitudinal polarization vector is given by $\epsilon^\mu(q,\lambda=0) = \frac{1}{\sqrt{Q^2}}(|\mathbf{q}|, \omega \hat{\mathbf{q}})$.  This has norm $\epsilon^\mu(q,\lambda=0) \epsilon_\mu(q,\lambda=0) = +1$.
 The Dirac blinear can then be decomposed as in Eq.~(\ref{decomp})  where the coefficients are $\Omega_\lambda =(\pm) \bar{u}(k') \gamma^\mu u(k) \epsilon_\mu(q,\lambda)$ with the $+$ sign taken for $\lambda = 0$ and $-$ for $\lambda = \pm$.
With this change, we can write the VCS amplitude as
 \begin{equation}
  \mathcal{M}_\text{VCS} =  \frac{i e}{q^2} \sum_\lambda \Omega_\lambda \epsilon_\mu(q,\lambda) \mathcal{T}^{\mu\nu}_\text{VCS} \epsilon'^*_\nu(q',\lambda').  \label{vcs_amp}
 \end{equation}

In Fig.~\ref{fig5}, the sub-diagram involving the two photons and Majorana fermion is identical to that of RCS, in Fig.~\ref{fig3}(a), except for the fact that the incident photon is virtual.  Because the incident photon can have a longitudinal mode, the contraction between $\epsilon^\mu(q,\lambda)$ and  $q^\mu$ is not necessarily zero; viz., $\epsilon \cdot q \ne 0$.  This can result in additional structures present in the low-energy expansion of the VCS Hamiltonian, but in the end, one should be able to recover the RCS Hamiltonian in the limit in which the incident photon becomes real.  With an appropriate linear combination of terms, this limiting process has been achieved by Gorchtein \cite{gorchtein_vcs}. 

To develop the LEX for the VCS Hamiltonian, Gorchtein finds it convenient to work in the Breit frame, which requires that the Majorana fermion's momentum pre- and post-interaction satisfy $\mathbf{p}'=-\mathbf{p}$. This choice  renders the same value for initial and final photon energies, $\omega$, which could not be achieved in, say, the center-of-mass frame because $q^2 \ne 0$.  Also, Gorchtein opts to work in the Coulomb gauge with spatial polarization vectors.  Recall, above in the Lorenz gauge, the longitudinal mode $\epsilon^\mu(q, \lambda=0)$ was time-like with $\epsilon \cdot q =0$.  Gauge invariance  allows us to redefine this vector so that it only has spatial components.  This does not impact the overall VCS amplitude by virtue of
the Ward identity, $q_\mu \mathcal{T}^{\mu \nu} =0$.   From $\epsilon^\mu(q,\lambda=0)$, we subtract $\frac{|\mathbf{q}|}{\omega \sqrt{Q^2}} q^\mu$ defining the longitudinal polarization vector as $\tilde\epsilon^\mu(q, \lambda=0) =  -\frac{\sqrt{Q^2}}{\omega}(0,  \hat{\mathbf{q}})$.  We note that this new vector, $\tilde\epsilon^\mu(q, \lambda=0)$, is no longer normalized nor orthogonal to $q^\mu$.  Going forward, we will use the three spatial polarization vectors $\{ \beps_\pm, \beps_0 \}$, dropping the tilde notation for the longitudinal mode.

With these conventions, Gorchtein determines the leading order contribution to the VCS Hamiltonian matrix element in a LEX is 
\al{
H_\text{VCS} \sim &     4\pi  \alpha_E(Q^2) \mathbf{E} \cdot \mathbf{E}' +  4\pi  \beta_M(Q^2) \mathbf{B} \cdot \mathbf{B}' \nonumber \\
&  +4\pi i  [\gamma_1(Q^2) -\gamma_2 (Q^2) -2 \gamma_4 (Q^2) ] \frac{1}{2}\{ \bsig \cdot [\mathbf{E} \times (\mathbf{q}' \times \mathbf{B}') ]-\bsig \cdot[ \mathbf{E}' \times( \mathbf{q} \times \mathbf{B}) ]\} \nonumber \\
& +4\pi i \gamma_2(Q^2) \frac{1}{2} \{ \bsig \cdot[\mathbf{q}' \times ( \mathbf{E}' \times \mathbf{B})] -  \bsig \cdot[\mathbf{q} \times ( \mathbf{E} \times \mathbf{B}')] \}    \nonumber\\
 &+ 4 \pi i \left[\frac{1}{2} \gamma_2(Q^2)  +\gamma_3(Q^2) + \gamma_4(Q^2)\right]\{(\bsig \cdot \mathbf{B}) (\mathbf{q} \cdot \mathbf{E}')- (\bsig \cdot \mathbf{B}') (\mathbf{q}' \cdot \mathbf{E})  \} \nonumber\\
 & - 4 \pi i \left[ \frac{1}{2} \gamma_2 (Q^2) + \gamma_4(Q^2)\right] (\bsig \cdot \boldsymbol{\Delta}) \{   \mathbf{B} \cdot \mathbf{E}' +  \mathbf{E} \cdot \mathbf{B}'\} \label{H_VCS_gorchtein}
,}
where the  spin-independent,  $\alpha_E(Q^2), \beta_M(Q^2)$, and spin-dependent, $\gamma_j(Q^2)$, generalized polarizabilities are functions of the initial photon's invariant ``mass", $Q^2$, and $\Delta^\mu := q-q' = p'-p$.  Gorchtein's definitions of $\gamma_j(Q^2)$ mirror their analogues in common RCS parameterizations, e.g., see Ref.~\cite{hhkk}.  To aid comparison with RCS, we will write this expression wholly in terms of the electromagnetic Fourier modes (still within the Breit frame)
\al{
H_\text{VCS} \sim &   (\omega^2 \alpha_E + (\mathbf{q} \cdot \mathbf{q}')\beta_M) \beps \cdot\beps'^*     -\beta_M (  {\beps}'^* \cdot { \mathbf{q}})  ( \beps \cdot \mathbf{q}' ) 
\nonumber \\
&+i \omega [\tfrac{1}{2}(|\mathbf{q}|^2+|\mathbf{q}'|^2 ) \gamma_1  -(\mathbf{q} \cdot \mathbf{q}' ) (\gamma_2 +2 \gamma_4 )  ] [ \bsig \cdot ({ \beps}'^{*} \times \beps)]   
\nonumber   \\
  &+i \omega \gamma_2  (\beps'^* \cdot \beps) [\bsig \cdot( \mathbf{q}' \times \mathbf{q})]  \nonumber \\
& + i \omega \gamma_4  \{       ( \beps  \cdot \mathbf{q}') [\bsig\cdot(\beps'^* \times \mathbf{q})]   -(\beps'^* \cdot \mathbf{q}) [\bsig\cdot(\beps \times \mathbf{q}')]    \} \nonumber \\
&+  i \omega  \gamma_3  \{ ({\beps}   \cdot   \mathbf{q}' )[\bsig\cdot (\beps'^*\times \mathbf{q}' )] - (\beps'^{*}    \cdot    \mathbf{q}) [\bsig \cdot  (\beps \times   \mathbf{q})]  \} \nonumber \\
&-i \omega \tfrac{1}{2} \gamma_1 (\beps \cdot \mathbf{q}) [\bsig \cdot (\beps'^* \times \mathbf{q})] + i\omega (\tfrac{1}{2} \gamma_2 + \gamma_4) (\beps \cdot \mathbf{q}) [\bsig \cdot (\beps'^*\times \mathbf{q}')],  \label{H_VCS_standard}
}
where we suppress the $Q^2$ dependence of the generalized polariziabilities.  Note that the last two terms in Eq.~(\ref{H_VCS_standard}) vanish for a transverse photon.  Additionally, if the incident photon were real, then in the Breit (and COM) frame, we would have $|\mathbf{q}|^2 = |\mathbf{q}'|^2 = \omega^2$.

\subsection{Generalized polarizabilities in the toy model}

Before computing the spin-flip transition rate mediated by the VCS vertex, we first compute the generalized polarizabilities for the toy model introduced in Sec.~\ref{toy}.  The Feynman diagrams that contribute to the generalized polarizabilities are the same as those in RCS; see
Fig.~\ref{fig4}.  The only modifications are that we set $q^2 = - Q^2$ and $q \cdot \epsilon \ne 0$.   We work in the Breit frame so that we may easily compare the resulting polarizabilities with those from RCS.
As before, we use {\tt Package-X 2.0} in {\tt Mathematica} to compute the diagrams \cite{patel} and carry out the low energy expansion. We treat both $q$ and $q'$ as small variables and carry all computations out to fifth order in these small quantities before truncating.  To simplify expressions, we also assume $M_\phi \gg m_\psi, m_\chi$, keeping at most terms that are $\mathcal{O}(M_\phi^{-4})$.  

For our toy model, we find the LEX of the VCS tensor in the Breit frame to be 
\al{
\mathcal{T}^{\mu\nu}_\text{VCS}\epsilon_\mu \epsilon'^*_\nu = & \frac{e^2 g^2}{(4\pi)^2}  \Bigg\{ \frac{m_\chi^2}{ M_\phi^4 } \left[
\left(
   \frac{4}{3}  \log \left(\frac{M_\phi^2}{m_\psi ^2}\right) -\frac{5}{3}  \right)  \omega^2  +
   \left( \frac{4}{3}  \log \left(\frac{M_\phi^2}{m_\psi ^2}\right) -\frac{13}{3}  
   \right) \mathbf{q} \cdot \mathbf{q}' 
    \right] (\beps \cdot \beps')   \nonumber  \\
& -\frac{ m_\chi^2}{ M_\phi^4} \left[ 
   \frac{4 }{3} \log \left(\frac{M_\phi^2}{m_\psi ^2}\right) -\frac{13}{3} \right](  {\beps}'^* \cdot { \mathbf{q}})  ( \beps \cdot \mathbf{q}' ) \nonumber\\
&+i \omega \left[   -\frac{2}{3}\frac{m_\chi}{ m_\psi^2 M_\phi^2}
    \left(\tfrac{1}{2} |\mathbf{q}|^2 + \tfrac{1}{2}|\mathbf{q}'|^2 - \mathbf{q} \cdot \mathbf{q}' \right)\right]  [ \bsig \cdot ({ \beps}'^{*} \times \beps)]   
\nonumber   \\
  &+i \omega \frac{m_\chi}{
  M_\phi^4 }   \left[  \frac{4}{3} \log\left(\frac{M_\phi^2}{m_\psi^2}\right) -3 \right] (\beps'^* \cdot \beps) [\bsig \cdot( \mathbf{q}' \times \mathbf{q})]  \nonumber \\
& - i \omega  \frac{1}{3} \frac{m_\chi}{ m_\psi^2 M_\phi^2 } \{       ( \beps  \cdot \mathbf{q}') [\bsig\cdot(\beps'^* \times \mathbf{q})]   -(\beps'^* \cdot \mathbf{q}) [\bsig\cdot(\beps \times \mathbf{q}')]    \} \nonumber \\
&+  i \omega   \frac{1}{3}\frac{m_\chi}{ m_\psi^2 M_\phi^2}  \{ ({\beps}   \cdot   \mathbf{q}' )[\bsig\cdot (\beps'^*\times \mathbf{q}' )] - (\beps'^{*}    \cdot    \mathbf{q}) [\bsig \cdot  (\beps \times   \mathbf{q})]  \} \nonumber \\
&-i \omega \frac{2}{3} \frac{m_\chi}{ m_\psi^2 M_\phi^2 }(\beps \cdot \mathbf{q}) [\bsig \cdot (\beps'^* \times \mathbf{q})] + i\omega \frac{2}{3} \frac{m_\chi}{m_\psi^2 M_\phi^2 }  (\beps \cdot \mathbf{q}) [\bsig \cdot (\beps'^*\times \mathbf{q}')].  \label{VCS_model_amp}
    \Bigg\} 
}
In the limit in which the incident photon becomes real, viz. $q^2=0$ and $\epsilon \cdot q=0$, we find the VCS amplitude is equivalent to the RCS amplitude, with the VCS generalized polarizabilities collapsing to the RCS ones calculated above; that is,  $\alpha_E(Q^2=0)$, $\beta_M(Q^2=0)$, and $\gamma_j(Q^2=0)$ correspond to their RCS counterparts in Eqs.~(\ref{alpha}-\ref{g4}).    However, the additional structures that are present for the longitudinal modes of the incident photon, those proportional to $\beps \cdot \mathbf{q}$, do not bear out the expectations from Gorchtein's LEX in Eq.~(\ref{H_VCS_standard}). From Eq.~(\ref{H_VCS_standard}), we expect the new VCS terms, relative to the RCS amplitude, to be $ i \omega ( -\tfrac{1}{2} \gamma_1   ) ( \beps \cdot \mathbf{q})   [\boldsymbol{\sigma} \cdot ( \beps'^* \times \mathbf{q})]   +   i \omega ( \tfrac{1}{2} \gamma_2  + \gamma_4 )      ( \beps \cdot \mathbf{q})  [ \boldsymbol{\sigma} \cdot ( \beps'^* \times \mathbf{q}') ] $.  Referring to the polarizabilities in our toy model, Eqs~(\ref{g1}-\ref{g4}), it is clear that we need to add two additional terms to Eq.~(\ref{H_VCS_gorchtein}): $ 4\pi i \gamma_5 \omega (\beps \cdot \mathbf{q}) [\bsig \cdot (\beps' \times \mathbf{q})]$ and $ 4\pi i \gamma_6 \omega  (\beps \cdot \mathbf{q}) [\bsig \cdot (\beps' \times \mathbf{q}')]$.  In our toy model, these new coefficients are found to be
$\gamma_5 \approx - \frac{e^2g^2}{(4\pi)^3} \frac{1}{2}   \frac{1}{M_\phi^2 m_\psi^2} $ and $\gamma_6 \approx  \frac{e^2g^2}{(4\pi)^3} \frac{1}{2}   \frac{1}{M_\phi^2 m_\psi^2} $.  

\subsection{Rate of  single-photon emission via VCS \label{vcs_rate_sec}}

With these generalized polarizabilities in hand, we can now compute the spin flip transition rate for the Majorana fermion in a background current $\mathbf{J} = J \hat{\mathbf{z}}$ that proceeds via the VCS Hamiltonian, Fig.~\ref{fig5}.  As before, we note that spin-down Majorana fermions, $\xi_-$, will have an energy of $\omega_0 = 2f_aJ$ greater than spin-up, $\xi_+$, counterparts. We, again, work in the rest frame of the Majorana fermion.  Because the generalized polarizabilities are frame dependent, they will pick up  corrections when boosting from the Breit frame to the particle's rest frame; in particular, there are terms that are $\mathcal{O}\left(\frac{Q^2}{m_\chi^2}\right)$. These corrections are subdominant, so we neglect them here. Additionally,  the VCS amplitude in the rest frame is modified because $\omega \ne \omega'$, which we accommodate.

The transition rate for the spin-flip process is give by
\begin{equation}
\Gamma_\text{VCS} = n_\Psi v_\Psi \sigma_\text{VCS},
\end{equation}
where $n_\Psi$ is the number density of  charged fermions incident upon the Majorana fermion, $v_\Psi$ their average speed, and $\sigma_\text{VCS}$ the total cross section for the VCS process.  To compute the total cross section, we sum over the final photon polarization states $\lambda'$ and integrate over the final available phase space for the charged fermion $\Psi$, real photon, and Majorana fermion.  

In our calculations, we will assume that the charged fermion is relativistic so that the particle's energy can be equated with the magnitude of its momentum, $E_\Psi \approx  |\mathbf{k}|$.  Likewise, for the charged fermion post-interaction, we will neglect the particle's mass. 
We will also assume the charged fermion to be unpolarized and sum over its final spin states.  
After integrating over the Majorana fermion's final momentum, $\boldsymbol{\Delta}$, the cross section becomes
\begin{equation}
\diff^6 \sigma_\text{VCS} = \frac{\diff^3 k' \, \diff^3 q'}{(4\pi)^5  v_\Psi m_\chi^2 |\mathbf{k}| |\mathbf{k}'| \omega'}  \delta({\textstyle \sum} E_i - E_f) \big|  \textstyle\sum_{\lambda'} \mathcal{M}_\text{VCS}\big|^2,
\end{equation}
with the VCS amplitude expressed previously in Eq.~(\ref{vcs_amp}), assuming now that we average over (sum over) the initial (final) charged fermion's spin states. We have made some additional kinematical simplifications. The Majorana fermion is taken to be at rest initially, and post-interaction we assume the impact on recoil to be minimal so that $p'^\mu \approx (m_\chi, \boldsymbol {\Delta})$.  Overall, momentum conservation in the interaction yields $\mathbf{k} - \mathbf{k}' = \mathbf{q} = \mathbf{q}' + \boldsymbol{\Delta}$ while energy conservation simplifies to $ |\mathbf{k}| =  |\mathbf{k}'| + \omega'  -\omega_0$.

The portion of the full VCS amplitude coming from the VCS tensor can be cast in terms of the generalized polarizabilities using our results from Eq.~(\ref{VCS_model_amp}).  Recall, the polarization vectors, $\beps$, in Eq.~(\ref{VCS_model_amp}) refer to the virtual photon and, in fact, we sum over the three-polarization states for the full VCS amplitude.   Working in the Coulomb gauge, this sum can be written as 
$ \sum_\lambda \Omega_\lambda \epsilon^\mu(q,\lambda) = \tilde{K}^\mu$ where $\tilde{K}^j = \bar{u}(k') \gamma^j u(k) -\bar{u}(k') \gamma^0 u(k) \omega^{-1} q^j$ and the temporal component vanishes by construction, $\tilde{K}^0=0$.  The leading order contribution to a spatial component of $\tilde{K}$ is given by the second term, $\tilde{K}^j \approx -\bar{u}(k') \gamma^0 u(k) \omega^{-1} q^j$; that is, the longitudinal polarization of the virtual photon is dominant in this process.   
Focusing only upon the spin-dependent processes for the Majorana fermion, the square of the amplitude becomes 
\al{
|\mathcal{M}_\text{VCS}|^2 = &\bigg| \frac{1}{2} \frac{e}{q^2} \sum_{\lambda, s, s'} \Omega_\lambda\epsilon_\mu(q,\lambda)  {\mathcal{T}^{\mu\nu}_\text{VCS}}_\text{spin} \epsilon'^*_\nu \bigg|^2 \\
 \approx & (|\mathbf{k}||\mathbf{k}'| + \mathbf{k}'\cdot \mathbf{k} ) \left(8\pi m_\chi\frac{e}{q^2}   \right)^2 \bigg| \sum_{\lambda'}\Big\{
 [  \gamma_5  |\mathbf{q}|^2+  \tfrac{1}{2} \gamma_1  |\mathbf{q}'|^2  -( \gamma_2 + \gamma_4)  (\mathbf{q} \cdot \mathbf{q}') ] [ \bsig \cdot ({ \beps}'^{*} \times \mathbf{q})]   
   \nonumber \\
& + ( \gamma_2 +\gamma_4)  (\beps'^* \cdot \mathbf{q}) [\bsig \cdot( \mathbf{q}' \times \mathbf{q})]     
\nonumber \\
&+[(\tfrac{1}{2} \gamma_2 + \gamma_4 +\gamma_6) |\mathbf{q}|^2+     \gamma_3 (  {\mathbf{q}}   \cdot   \mathbf{q}' )][\bsig\cdot (\beps'^*\times \mathbf{q}' )]   \Big\}  \bigg|^2 ,
}
where $s$ $(s')$ indicate the initial (final) spin states of the charged fermion, and we neglect the charged fermion's mass.

The integration over the phase space of the of the final photon is routine, shifting over to spherical coordinates for the phase space variables.  In terms of integration limits, the minimum momentum transfer to the Majorana fermion must be sufficient to produce a real photon of energy $\omega_0$ so we require $|\mathbf{q}|\ge \omega_0$; this provides an upper limit for the final-state photon's energy, $|\mathbf{k}'| \le |\mathbf{k}| - \omega_0$.   Finally, we assume that the momentum, $\mathbf{k}$, of the initial charged fermion is randomly oriented as it would be in a plasma. In our final expression for the cross section, we average over all possible directions $\mathbf{k}$.
 In the end, we find the leading order contribution to the cross section is
\al{
\sigma = & \frac{e^2}{\pi}\bigg\{   |\mathbf{k}|^6 \bigg( \frac{1}{528}  \gamma_1^2  + \frac{1}{180}  \gamma_2^2     + \frac{1}{280} \gamma_3^2       +    \frac{1}{166} \gamma_4^2     + \frac{43}{540} \gamma_5^2    +   \frac{1}{45} \gamma_6^2 \\
&+ \frac{1}{264}   \gamma_1 \gamma_3  +  \frac{1}{62}   \gamma_1 \gamma_5   +   \frac{1}{166}  \gamma_2 \gamma_4  -\frac{1}{97}   \gamma_2 \gamma_6  +  \frac{1}{62}   \gamma_3 \gamma_5 +  \frac{1}{83}  \gamma_4 \gamma_6  \bigg)  \\
& + |\mathbf{k}|^5 \omega_0 \bigg(   \frac{1}{64}  \gamma_1^2 
  +    \frac{1}{30}   \gamma_2^2   +   \frac{3}{101}  \gamma_3^2 +    \frac{6}{163}  \gamma_4^2      +    \frac{31}{90}  \gamma_5^2    +   \frac{2}{15}  \gamma_6^2  \\
  &  +  \frac{5}{159}  \gamma_1 \gamma_3  
   +   \frac{8}{83} \gamma_1 \gamma_5    +     \frac{6}{163}    \gamma_2 \gamma_4  -\frac{5}{84} \gamma_2 \gamma_6  +    \frac{8}{83}   \gamma_3 \gamma_5   +   \frac{9}{122}  \gamma_4 \gamma_6 \bigg)\Bigg\}  \label{sigma}
}
neglecting terms that are higher order in $\omega_0$\footnote{Note, in Eq.~\ref{sigma} the coefficients to each term were evaluated via numerical integration and then rationalized to a precision of one part in ten thousand.}.  For the toy model, we again define $\gamma := \gamma_3 = \frac{e^2g^2}{(4\pi)^3} \frac{1}{6}\frac{1}{M_\phi^2m_\psi^2}$, and then we can write the cross section in terms of this parameter. 
The resulting rate of spin-transition, from antialigned to aligned,  through the VCS channel is
\begin{equation}
\Gamma_\text{VCS} = \alpha \gamma^2   n_\Psi  |\mathbf{k}|^5 \left( 3.8  |\mathbf{k}|  + 17.7 \omega_0 \right),
\end{equation}
where $\alpha$ is the fine structure constant.  

One important distinction, relative to the two-photon spontaneous emission process considered previously, is that in the VCS process the spin of the Majorana fermion can flip from aligned with the external current to antialigned.  That is, if the virtual photon from the charged fermion, $\Psi$,  is sufficiently energetic, namely $\omega > \omega_0$, then the Majorana fermion can be converted into the more energetic antialigned state through single photon emission.  This process competes with the one in which the Majorana fermion's spin flips to the lower energy state.  For a collection of unpolarized Majorana fermions, it is the difference between these two rates that determines the rate at which the system can be polarized through the VCS process.  After computing the aligned-to-antialigned transition rate, we find the difference in spin-flip rates to be
\begin{equation}
\Delta \Gamma_\text{VCS}  = 35.5  \alpha \gamma^2   n_\Psi   |\mathbf{k}|^5\omega_0, \label{dgamma_VCS}
\end{equation}
to leading order for the toy model. 

\section{ Discussion  }

Because dark matter interactions are highly suppressed, we anticipated at the outset of this calculation that the ability to align with a background current the spins of  a collection of DM Majorana fermions, via either two-photon spontaneous emission or single-photon emission through VCS, is nearly impossible in all but the most extreme environments.  Present day, reasonable parameter estimates for the size of the anapole moment and polarizabilities show that the irreversible spin-flip transitions discussed above are essentially forbidden in terrestrial experiments and stellar environments. But, in the early universe, the plasma density is so large that the differential spin-flip rates for the VCS process could be significant if sufficiently large currents exist. 

In the very early universe when temperatures are much greater than the DM mass, $T\gg m_\chi$, reversible spin-flip interactions can assist in polarizing a DM medium in the presence of a background current.  Because DM is in thermal equilibrium with the thermal bath, one would expect that the distribution of DM states would follow the Boltzmann distribution with the ratio of states aligned with a background current to those antialigned given by $n(\uparrow)/n(\downarrow) = \exp(\omega_0/T)$.
As the universe cools, $T \lesssim m_\chi$, DM becomes non-relativistic, and because it interacts so weakly, DM eventually decouples from the thermal bath. At freeze out, the DM system is fully decoupled, and the DM's comoving number density essentially becomes constant.  After freeze out,  irreversible processes like two-photon emission and VCS are the only available mechanisms for polarization.  We examine these here. 

The freeze out temperature, $T_f$, and relic DM density are largely determined by the DM annihilation cross section.  For a large annihilation cross section, the DM stays in thermal equilibrium until later times (that is, lower temperatures) resulting in  greater DM annihilation, reducing the ultimate relic density of the particle. Conversely, for smaller annihilation cross sections, freeze occurs at a higher temperature, and the relic DM density is greater.  To precisely determine the temperature at which thermal DM  begins to drop out of equilibrium one must solve the Boltzmann equation, but reliable estimates of the freeze out temperature and relic density, accurate to a few percent, have been developed.

  Before estimating the freeze out temperature,  it is first useful to expand the annihilation cross section as a power series in velocity. Because DM is non-relativistic at freeze out, the lower velocity modes will be the most significant annihilation channels.  This expansion essentially encodes the temperature dependence of the annihilation mode given that   $\langle |v|\rangle \sim T^\frac{1}{2}$ for non-relativistic particles.  In many models, a single annihilation mode dominates. In this case, we can parametrize the thermally averaged annihilation cross section as $\langle \sigma_\text{ann} |v| \rangle = \sigma_0 x^{-n}$ where we employ the dimensional parameter $x := \frac{m_\chi}{T}$ and $\sigma_0$ has no explicit temperature dependence  \cite{kolb_turner}.  The $s$-wave annihilation mode corresponds to $n=0$, $p$-wave to $n=1$, and so on.  Assuming a dominant annihilation mode,  one may estimate that freeze out occurs when
\begin{equation}
x_f  \simeq \log \left[  0.076\,\, g_*^{-\frac{1}{2}}(n+1) M_\text{Pl} m_\chi  \sigma_0   \right] -\left(n+\frac{1}{2}\right)  \log[ \log[  0.076\,\, g_*^{-\frac{1}{2}}(n+1) M_\text{Pl} m_\chi \sigma_0 ]],  \label{xf}
\end{equation}
where $g_*$ represents the relativistic degrees of freedom at freeze out and $M_\text{Pl}$ is the Planck mass \cite{Scherrer:1985zt,kolb_turner}.   Once the freeze out temperature is determined, one may estimate the relative DM relic density present today
\al{
\Omega_\text{DM}  =& \frac{3.79\,  s_0}{\rho_\text{crit}  g_*^{\frac{1}{2}} M_\text{Pl}} \frac{(n+1)x_f^{n+1} }{ \sigma_{0}}, \label{omegadm}
}
where $s_0$ represents the universe's present entropy density and $\rho_\text{crit}$ is the critical energy density  \cite{Scherrer:1985zt,kolb_turner}. We use the values for the necessary astrophysical data found in Ref.~\cite{pdg2020}.

To determine the freeze out temperature for the toy model, we must first examine its  annihilation modes. 
The $s$-wave annihilation of DM into two photons proceeds via the polarizability terms in the LEX of the interaction Hamiltonian, Eq.~(\ref{H_pol}).  In terms of kinematics, the initial momenta of the DM particles are $\mathbf{p} \sim \mathbf{p}'\sim 0$. Momentum conservation results in two back-to-back final state real photons with momenta $q = m_\chi(1, \hat{\mathbf{q}})$ and $q' = m_\chi(1, -\hat{\mathbf{q}})$; additionally, the photons are transverse, $\mathbf{q} \cdot \beps = \mathbf{q}' \cdot \beps' = 0$.  Given these kinematic constraints, an examination of the RCS Hamiltonian matrix elements in Eq.~(\ref{rcs_ham}) suggests that the only surviving term in the $s$-wave annihilation amplitude has the form  $\mathcal{M}_s \sim  m_\chi^4 [     \gamma_1+ \gamma_2 + 2 \gamma_4  ] [ \bsig \cdot ({ \beps}'^{*} \times \beps)]  \sim g^2 e^2 \frac{m_\chi^4}{M_\phi^4} [ \bsig \cdot ({ \beps}'^{*} \times \beps)] $, using the polarizabilities in Eqs.~(\ref{g1}-\ref{g4}).  A more refined treatment of the mass scales involved in the process result in amplitudes with different mass dependencies \cite{anapole_2photon}, but if the loop masses dominate the DM mass, $M_\phi, m_\psi \gg m_\chi$, then indeed the $s$-wave annihilation amplitude scales as $\mathcal{M}_s \sim \frac{m_\chi^4}{M^4}$ where $M \sim M_\phi, m_\psi$.   With this amplitude, the annihilation cross section for the $s$-wave process scales as $\sigma_s \sim \frac{m_\chi^6}{M^8}$.
We will see that this mass dependence suppresses the $s$-wave annihilation mode relative to the $p$-wave mode.  We also note that even if we have the alternate mass hierarchy $M_\phi \gg m_\chi \gg m_\psi$, $p$-wave annihilation still dominates the $s$-process for large ($\mathcal{O}$ (100 GeVs)) DM masses because the $p$-wave channel has a multitude of kinematically available final states \cite{anapole_2photon}.
 
In $p$-wave annihilation, the non-relativistic DM particles couple  via the anapole moment to a virtual photon which pair produces charged particles like an electron and positron. This annihilation cross section in the center of mass frame has been worked out previously \cite{anapole_dm1}
\begin{equation}
\sigma_p v_\text{rel} = \frac{2}{3} \alpha f_a^2 m_\chi^2 v_\text{rel}^2.
\end{equation}
  The mass dependence of the anapole moment in our toy model is $f_a \sim \frac{1}{M_\phi^2}$, so the $p$-wave annihilation cross section scales as $\sigma_p \sim \frac{m_\chi^2}{M_\phi^4}$. This mode dominates the $s$-wave cross section, so we assume annihilation proceeds predominantly through the $p$-wave mode. Expressing this is terms of the dimensionless factor $x$, the thermally averaged cross section is
\begin{equation}
\langle \sigma_p |v_\text{rel}| \rangle =  \frac{2}{3} \alpha N_f f_a^2 m_\chi^2 \langle |v_\text{rel}|^2 \rangle= 
\sigma_0 x^{-1}  
\end{equation}
with $\sigma_0 =4 \alpha N_f f_a^2 m_\chi^2$.
As in Ref.~\cite{anapole_dm1}, the factor $N_f$ accounts for the sum over all the kinematically available final states weighted by the square of the particles' charges.   For instance, an electron-positron pair contribute a term of 1 to the sum, whereas an up-antiup pair contribute a term of $\left(\frac{2}{3}\right)^2\times 3$ to the sum, where the factor of 3 accounts for the different color states. Because the $p$-wave annihilation mode is much larger than the $s$-wave mode for reasonable model parameters,  we set $n=1$ in Eqs.~(\ref{xf}, \ref{omegadm}).

After freeze out, the only mechanism by which polarization of a DM medium can be achieved is via irreversible processes like those discussed above. But this can only happen if the spin-flip rate aligning the anapole moment with the background current is much greater than the universe's expansion rate.  When DM decouples, the universe is in a radiation dominated era, and the Hubble parameter  is given by  \cite{kolb_turner}
\begin{equation}
H = 1.66 g_\star^\frac{1}{2} \frac{T^2}{M_\text{Pl}}.
\end{equation}
In order for the DM to achieve some degree of polarization, we require that the irreversible spin-flip rate is much greater than the Hubble parameter around freeze out; $\Gamma \gg H$. 

 Let us consider specific model parameters to see if this inequality can be achieved.
Collider experiments place meaningful constraints on the size of possible anapole moments for particles with masses on the order of hundreds of GeVs.  Using an effective-field-theory approach, a recent study, Ref.~\cite{alves},  finds that the high-luminosity LHC runs constrain the anapole moment for a 500 GeV particle to be smaller than $f_a = 2\times 10^{-6} \mu_N$ fm.  Given this value for the anapole moment, we can use Eq.~(\ref{omegadm}) to directly relate the DM's anapole moment to the freeze-out temperature, if we assume the entirety of the relic DM is due to this Majorana fermion.    Taking $f_a = 2\times 10^{-6} \mu_N$ fm, we find that  freeze out occurs at $T_f = 17$ GeV.   Then, using Eq.~(\ref{xf}), this freeze out temperature and anapole moment correspond to a DM mass of $m_\chi = 419$ GeV.  This is reasonably consistent with the constraints from Ref.~\cite{alves}.   We will also need to estimate the size of the polarizabilties.  To do so requires an assumption about the mass of the charged fermion $\psi$ to which the Majorana fermion couples.  We set $m_\psi = 100$ GeV, which yields  $\gamma := \gamma_3 \sim 6.6 \times 10^{-13}$ GeV$^{-4}$, neglecting an $\mathcal{O}(1)$ logarithm, $\log(M_\phi/m_\psi)$.

Turning our attention to spin-flip processes, we now examine whether spontaneous two-photon emission is viable at freeze out for DM with this particular anapole moment.  The rate depends heavily on the energy difference between the two spin states, $\omega_0 = 2 f_a J$, dictated by the background current density.   We will assume the current is due to some net drift velocity of charged particles in the relativistic thermal plasma, $J = e n_q v_\text{drift}$.  The factor $n_q$ is the number density of charge carriers summed over all the charged relativistic species weighted by the absolute value of their charge; e.g., electrons are weighted by $1$ while up quarks have a weight of $\frac{2}{3}$. 

 Below the freeze out temperature of $17$ GeV, the charged relativistic species are exclusively fermionic, and the number density for a single relativistic species of fermions is 
\begin{equation}
n = \frac{3}{4} \frac{\zeta(3)}{\pi^2} g T^3
\end{equation}
where $\zeta$ is the Riemann zeta function and $g$ is the degrees of freedom for the particular  species   \cite{kolb_turner}. Around the freeze out temperature, we have $J = 1.1\, v_\text{drift} \, T^3$ which yields an energy difference of  $\omega_0=18$ MeV, taking $v_\text{drift}=1$ to represent an upper limit on the drift speed.   From Eq.~(\ref{spont_rate}), we find that the two-photon spontaneous emission rate at freeze out is $\Gamma_\text{spont} \le  4.6 \times 10^{-19}$ s$^{-1}$, which corresponds to a transition time  that is on the order of the Hubble time.  Comparing this to the universe's expansion rate at $T_f = 17$ GeV,  we find  $\frac{\Gamma_\text{spont}}{H} \sim 8.3 \times 10^{-28}$.  Thus, the rate of spontaneous two-photon emission is negligible around DM freeze out for these model parameters.

We now consider the differential spin-flip rate for the VCS process.   In Sec.~\ref{vcs_rate_sec}, we assumed that VCS was effectuated by a single charged fermion of momentum $\mathbf{k}$, though in the early universe VCS would be due to an incoherent sum of interactions with various fermion species in the relativistic plasma.  To incorporate in the differential spin-flip rate, Eq.~(\ref{dgamma_VCS}), several flavors of fermions with differing charges, we substitute for $n_\Psi$ a sum over the relativistic flavors of fermions, at the given temperature, with each fermion's contribution weighted by the square of its charge. Additionally, the fermion momentum is given by the thermal average \cite{kolb_turner}
\begin{equation}
\langle |\mathbf{k}| \rangle = \frac{7\pi^4}{180\, \zeta(3)} T.
\end{equation}
Evaluating the degrees of freedom at $17$ GeV, we compare the differential spin-flip rate to the Hubble parameter
\begin{equation}
\frac{\Delta \Gamma_\text{VCS}}{H} = 28 M_\text{Pl} f_a \gamma^2 T^9 v_\text{drift}.
\end{equation}
At freeze out for the model parameters under consideration, we find that the differential spin-flip rate is sufficiently large if $v_\text{drift} \gg 0.03$.

The existence of such currents in the early universe would break the homogeneity assumed in the cosmological principle, but  inhomogeneities must occur to seed the formation of galaxies.  In the literature, we find that the  production of electric currents in the early universe has been discussed in the inflationary period \cite{PhysRevD.37.2743}, around the electroweak phase transition \cite{PhysRevD.55.4582,PhysRevD.53.662}, and around the QCD phase transition \cite{QCDtrans_currents, PhysRevD.50.2421,PhysRevD.55.4582,PhysRevD.77.043529}.  These currents are expected to dampen rather quickly, compared to the Hubble time \cite{siegel_fry}, but their ability to locally polarize DM does not rely on long time scales.  

 It is beyond the scope of this present paper to determine if sufficient currents exist to polarize a DM medium at freeze out, but if they do exist, there are implications for DM indirect detection experiments.    Supposing a DM medium were to be partially polarized at freeze out by an electric current, the DM polarization in the local medium will  be locked in place because subsequent interactions with the thermal bath will be too feeble to disrupt it. The consequence of such polarized regions, present day, is the suppression  of the rate of $s$-wave annihilation which, by definition, requires vanishing total angular momentum.    For a Majorana fermion,  $s$-wave annihilation results in two final state photons, coupled via the fermion's polarizabilities, and an indirect DM detection experiment could look for high-energy photons as the annihilation products of DM. If no photons were observed, the experiment places an upper bound on the annihilation cross section as a function of DM  mass.  However, if the detector were within a region of partially polarized DM, the established bound on the $s$-wave annihilation cross section would be an overestimate if the DM medium were assumed to be unpolarized.   

\section{conclusion}

In this paper, we studied the ability of background currents to polarize Majorana fermions through irreversible processes involving photon emission.  If a Majorana fermion has a non-vanishing anapole moment, then a background current results in an energy difference between states with spins aligned and antialigned with the current.  Because the other static electromagnetic properties of a Majarona fermion necessarily vanish, the ability of a Majorana fermion to flip its spin in the presence of such a background current requires higher order interactions. We focused upon two possibilities.  The first involved spontaneous two-photon emission which proceeds via the particle's polarizability.  The second involved emission of a single photon as would occur in virtual Compoton scattering; this proceeds by virtue of the particle's generalized polarizabilities.

 Both processes involve spin-dependent effective couplings  that carry mass dimension [M]$^{-4}$ with a GeV-mass scale.
 As such, they are highly suppressed making the transition rate from the state in which the anapole moment is antialigned with the current to the one in which it is aligned vanishingly small in terrestrial experiments or stellar environments.  However, if DM is constituted by massive Majorana fermions, the early universe allows for the possibility that these spin-flip interactions can occur at an appreciable rate.  We focused on the time at which DM decouples from the thermal bath.   By considering reasonable parameter values for a thermal DM model, we find that spontaneous two-photon emission is negligible at this time, but single-photon emission via VCS can be appreciable.  This has consequences for present-day indirect DM searches.  If a region of DM were polarized by a background current at freeze out, then the persistent polarization would suppress $s$-wave annihilation, potentially confounding limits on the annihilation cross section set by the indirect detection experiment.
 
Our focus in this paper has been on irreversible interactions that flip the spin of a Majorana fermion in a current.  If the Majorana fermions are not in thermal equilibrium, then there are no other mechanisms by which a collection of these particles can be polarized.
However, if we consider an earlier time in the universe before freeze out,  Majorana fermions interact sufficiently to thermalize, and reversible interactions  can facilitate the spin-flip transition. Because these interactions occur via the anapole coupling and not a polarizability, it is likely that the rate of interaction is dramatically higher.
The situation is akin to what occurs in NMR physics.  The transition rate for the proton's magnetic dipole moment to align with a $\sim 10$ T magnetic field via spontaneous emission of a single photon, an irreversible process, was found to be $10^{-22}$ s$^{-1}$ \cite{purcell_spontaneous_1946}.  But NMR relaxation times, due to reversible spin-spin interactions or spin-lattice interactions, are on the order of seconds.  If our Majorana fermions can interact with the thermal bath, then polarization might be more easily achieved.  We will pursue this line of thought in future work.

\section{ACKNOWLEDGMENTS}
DCL thanks the Kavli Institute for Theoretical Physics for its hospitality during the completion of this work. 
This research was supported in part by the National Science Foundation under Grant No. NSF PHY-1748958.

\bibliography{biblio}

\begin{thebibliography}{57}
\expandafter\ifx\csname natexlab\endcsname\relax\def\natexlab#1{#1}\fi
\expandafter\ifx\csname bibnamefont\endcsname\relax
  \def\bibnamefont#1{#1}\fi
\expandafter\ifx\csname bibfnamefont\endcsname\relax
  \def\bibfnamefont#1{#1}\fi
\expandafter\ifx\csname citenamefont\endcsname\relax
  \def\citenamefont#1{#1}\fi
\expandafter\ifx\csname url\endcsname\relax
  \def\url#1{\texttt{#1}}\fi
\expandafter\ifx\csname urlprefix\endcsname\relax\def\urlprefix{URL }\fi
\providecommand{\bibinfo}[2]{#2}
\providecommand{\eprint}[2][]{\url{#2}}

\bibitem[{\citenamefont{Kayser}(1982)}]{bk82}
\bibinfo{author}{\bibfnamefont{B.}~\bibnamefont{Kayser}},
  \bibinfo{journal}{Phys. Rev.} \textbf{\bibinfo{volume}{D26}},
  \bibinfo{pages}{1662} (\bibinfo{year}{1982}).

\bibitem[{\citenamefont{Nieves}(1982)}]{nieves}
\bibinfo{author}{\bibfnamefont{J.~F.} \bibnamefont{Nieves}},
  \bibinfo{journal}{Phys. Rev.} \textbf{\bibinfo{volume}{D26}},
  \bibinfo{pages}{3152} (\bibinfo{year}{1982}).

\bibitem[{\citenamefont{Gray et~al.}(2010)\citenamefont{Gray, Karl, and
  Novikov}}]{gray_karl_novikov}
\bibinfo{author}{\bibfnamefont{C.~G.} \bibnamefont{Gray}},
  \bibinfo{author}{\bibfnamefont{G.}~\bibnamefont{Karl}}, \bibnamefont{and}
  \bibinfo{author}{\bibfnamefont{V.~A.} \bibnamefont{Novikov}},
  \bibinfo{journal}{Am. J. Phys.} \textbf{\bibinfo{volume}{78}},
  \bibinfo{pages}{936} (\bibinfo{year}{2010}).

\bibitem[{\citenamefont{Zeldovich}(1957)}]{zeldovich}
\bibinfo{author}{\bibfnamefont{{\relax Ya}.~B.} \bibnamefont{Zeldovich}},
  \bibinfo{journal}{Sov. Phys. JETP} \textbf{\bibinfo{volume}{6}},
  \bibinfo{pages}{1184} (\bibinfo{year}{1957}), \bibinfo{note}{[Zh. Eksp. Teor.
  Fiz. {\bf 33},1531 (1957)]}.

\bibitem[{\citenamefont{Radescu}(1985)}]{radescu}
\bibinfo{author}{\bibfnamefont{E.~E.} \bibnamefont{Radescu}},
  \bibinfo{journal}{Phys. Rev.} \textbf{\bibinfo{volume}{D32}},
  \bibinfo{pages}{1266} (\bibinfo{year}{1985}).

\bibitem[{\citenamefont{Prange}(1958)}]{prange}
\bibinfo{author}{\bibfnamefont{R.~E.} \bibnamefont{Prange}},
  \bibinfo{journal}{Phys. Rev.} \textbf{\bibinfo{volume}{110}},
  \bibinfo{pages}{240} (\bibinfo{year}{1958}).

\bibitem[{\citenamefont{Chen et~al.}(2001)\citenamefont{Chen, Cohen, and
  Kao}}]{chen}
\bibinfo{author}{\bibfnamefont{J.-W.} \bibnamefont{Chen}},
  \bibinfo{author}{\bibfnamefont{T.~D.} \bibnamefont{Cohen}}, \bibnamefont{and}
  \bibinfo{author}{\bibfnamefont{C.~W.} \bibnamefont{Kao}},
  \bibinfo{journal}{Phys. Rev.} \textbf{\bibinfo{volume}{C64}},
  \bibinfo{pages}{055206} (\bibinfo{year}{2001}).

\bibitem[{\citenamefont{Gorchtein}(2008)}]{gorchtein}
\bibinfo{author}{\bibfnamefont{M.}~\bibnamefont{Gorchtein}},
  \bibinfo{journal}{Phys. Rev.} \textbf{\bibinfo{volume}{C77}},
  \bibinfo{pages}{065501} (\bibinfo{year}{2008}).

\bibitem[{\citenamefont{Latimer}(2016)}]{maj_2photon}
\bibinfo{author}{\bibfnamefont{D.~C.} \bibnamefont{Latimer}},
  \bibinfo{journal}{Phys. Rev.} \textbf{\bibinfo{volume}{D94}},
  \bibinfo{pages}{093010} (\bibinfo{year}{2016}).

\bibitem[{\citenamefont{Ragusa}(1993)}]{ragusa}
\bibinfo{author}{\bibfnamefont{S.}~\bibnamefont{Ragusa}},
  \bibinfo{journal}{Phys. Rev.} \textbf{\bibinfo{volume}{D47}},
  \bibinfo{pages}{3757} (\bibinfo{year}{1993}).

\bibitem[{\citenamefont{Ragusa}(1994)}]{ragusa2}
\bibinfo{author}{\bibfnamefont{S.}~\bibnamefont{Ragusa}},
  \bibinfo{journal}{Phys. Rev. D} \textbf{\bibinfo{volume}{49}},
  \bibinfo{pages}{3157} (\bibinfo{year}{1994}).

\bibitem[{\citenamefont{Jacob and Mathews}(1960)}]{jacob_mathews}
\bibinfo{author}{\bibfnamefont{M.}~\bibnamefont{Jacob}} \bibnamefont{and}
  \bibinfo{author}{\bibfnamefont{J.}~\bibnamefont{Mathews}},
  \bibinfo{journal}{Phys. Rev.} \textbf{\bibinfo{volume}{117}},
  \bibinfo{pages}{854} (\bibinfo{year}{1960}).

\bibitem[{\citenamefont{Hearn and Leader}(1962)}]{hearn}
\bibinfo{author}{\bibfnamefont{A.~C.} \bibnamefont{Hearn}} \bibnamefont{and}
  \bibinfo{author}{\bibfnamefont{E.}~\bibnamefont{Leader}},
  \bibinfo{journal}{Phys. Rev.} \textbf{\bibinfo{volume}{126}},
  \bibinfo{pages}{789} (\bibinfo{year}{1962}).

\bibitem[{\citenamefont{Bardeen and Tung}(1968)}]{bardeen}
\bibinfo{author}{\bibfnamefont{W.~A.} \bibnamefont{Bardeen}} \bibnamefont{and}
  \bibinfo{author}{\bibfnamefont{W.~K.} \bibnamefont{Tung}},
  \bibinfo{journal}{Phys. Rev.} \textbf{\bibinfo{volume}{173}},
  \bibinfo{pages}{1423} (\bibinfo{year}{1968}), \bibinfo{note}{[Erratum: Phys.
  Rev. {\bf D4}, 3229 (1971)]}.

\bibitem[{\citenamefont{Bernard et~al.}(1991)\citenamefont{Bernard, Kaiser, and
  Meissner}}]{Bernard:1991rq}
\bibinfo{author}{\bibfnamefont{V.}~\bibnamefont{Bernard}},
  \bibinfo{author}{\bibfnamefont{N.}~\bibnamefont{Kaiser}}, \bibnamefont{and}
  \bibinfo{author}{\bibfnamefont{U.~G.} \bibnamefont{Meissner}},
  \bibinfo{journal}{Phys. Rev. Lett.} \textbf{\bibinfo{volume}{67}},
  \bibinfo{pages}{1515} (\bibinfo{year}{1991}).

\bibitem[{\citenamefont{Hemmert et~al.}(1998)\citenamefont{Hemmert, Holstein,
  Kambor, and Knochlein}}]{hhkk}
\bibinfo{author}{\bibfnamefont{T.~R.} \bibnamefont{Hemmert}},
  \bibinfo{author}{\bibfnamefont{B.~R.} \bibnamefont{Holstein}},
  \bibinfo{author}{\bibfnamefont{J.}~\bibnamefont{Kambor}}, \bibnamefont{and}
  \bibinfo{author}{\bibfnamefont{G.}~\bibnamefont{Knochlein}},
  \bibinfo{journal}{Phys. Rev.} \textbf{\bibinfo{volume}{D57}},
  \bibinfo{pages}{5746} (\bibinfo{year}{1998}).

\bibitem[{\citenamefont{Babusci et~al.}(1998)\citenamefont{Babusci, Giordano,
  L'vov, Matone, and Nathan}}]{babusci}
\bibinfo{author}{\bibfnamefont{D.}~\bibnamefont{Babusci}},
  \bibinfo{author}{\bibfnamefont{G.}~\bibnamefont{Giordano}},
  \bibinfo{author}{\bibfnamefont{A.~I.} \bibnamefont{L'vov}},
  \bibinfo{author}{\bibfnamefont{G.}~\bibnamefont{Matone}}, \bibnamefont{and}
  \bibinfo{author}{\bibfnamefont{A.~M.} \bibnamefont{Nathan}},
  \bibinfo{journal}{Phys. Rev.} \textbf{\bibinfo{volume}{C58}},
  \bibinfo{pages}{1013} (\bibinfo{year}{1998}).

\bibitem[{\citenamefont{Holstein et~al.}(2000)\citenamefont{Holstein, Drechsel,
  Pasquini, and Vanderhaeghen}}]{holstein_hopol}
\bibinfo{author}{\bibfnamefont{B.~R.} \bibnamefont{Holstein}},
  \bibinfo{author}{\bibfnamefont{D.}~\bibnamefont{Drechsel}},
  \bibinfo{author}{\bibfnamefont{B.}~\bibnamefont{Pasquini}}, \bibnamefont{and}
  \bibinfo{author}{\bibfnamefont{M.}~\bibnamefont{Vanderhaeghen}},
  \bibinfo{journal}{Phys. Rev.} \textbf{\bibinfo{volume}{C61}},
  \bibinfo{pages}{034316} (\bibinfo{year}{2000}).

\bibitem[{\citenamefont{Bedaque and Savage}(2000)}]{bedaque}
\bibinfo{author}{\bibfnamefont{P.~F.} \bibnamefont{Bedaque}} \bibnamefont{and}
  \bibinfo{author}{\bibfnamefont{M.~J.} \bibnamefont{Savage}},
  \bibinfo{journal}{Phys. Rev.} \textbf{\bibinfo{volume}{C62}},
  \bibinfo{pages}{018501} (\bibinfo{year}{2000}).

\bibitem[{\citenamefont{Bernard et~al.}(2003)\citenamefont{Bernard, Hemmert,
  and Meissner}}]{Bernard:2002pw}
\bibinfo{author}{\bibfnamefont{V.}~\bibnamefont{Bernard}},
  \bibinfo{author}{\bibfnamefont{T.~R.} \bibnamefont{Hemmert}},
  \bibnamefont{and} \bibinfo{author}{\bibfnamefont{U.-G.}
  \bibnamefont{Meissner}}, \bibinfo{journal}{Phys. Rev.}
  \textbf{\bibinfo{volume}{D67}}, \bibinfo{pages}{076008}
  (\bibinfo{year}{2003}).

\bibitem[{\citenamefont{Hildebrandt et~al.}(2004)\citenamefont{Hildebrandt,
  Griesshammer, and Hemmert}}]{Hildebrandt:2003md}
\bibinfo{author}{\bibfnamefont{R.~P.} \bibnamefont{Hildebrandt}},
  \bibinfo{author}{\bibfnamefont{H.~W.} \bibnamefont{Griesshammer}},
  \bibnamefont{and} \bibinfo{author}{\bibfnamefont{T.~R.}
  \bibnamefont{Hemmert}}, \bibinfo{journal}{Eur. Phys. J.}
  \textbf{\bibinfo{volume}{A20}}, \bibinfo{pages}{329} (\bibinfo{year}{2004}).

\bibitem[{\citenamefont{Holstein et~al.}(2005)\citenamefont{Holstein,
  Pascalutsa, and Vanderhaeghen}}]{holstein_sumrules}
\bibinfo{author}{\bibfnamefont{B.~R.} \bibnamefont{Holstein}},
  \bibinfo{author}{\bibfnamefont{V.}~\bibnamefont{Pascalutsa}},
  \bibnamefont{and}
  \bibinfo{author}{\bibfnamefont{M.}~\bibnamefont{Vanderhaeghen}},
  \bibinfo{journal}{Phys. Rev.} \textbf{\bibinfo{volume}{D72}},
  \bibinfo{pages}{094014} (\bibinfo{year}{2005}).

\bibitem[{\citenamefont{Guichon et~al.}(1995)\citenamefont{Guichon, Liu, and
  Thomas}}]{guichon}
\bibinfo{author}{\bibfnamefont{P.~A.~M.} \bibnamefont{Guichon}},
  \bibinfo{author}{\bibfnamefont{G.~Q.} \bibnamefont{Liu}}, \bibnamefont{and}
  \bibinfo{author}{\bibfnamefont{A.~W.} \bibnamefont{Thomas}},
  \bibinfo{journal}{Nucl. Phys. A} \textbf{\bibinfo{volume}{591}},
  \bibinfo{pages}{606} (\bibinfo{year}{1995}).

\bibitem[{\citenamefont{Vanderhaeghen}(1996)}]{vanderhaeghen}
\bibinfo{author}{\bibfnamefont{M.}~\bibnamefont{Vanderhaeghen}},
  \bibinfo{journal}{Phys. Lett. B} \textbf{\bibinfo{volume}{368}},
  \bibinfo{pages}{13} (\bibinfo{year}{1996}).

\bibitem[{\citenamefont{Scherer et~al.}(1996)\citenamefont{Scherer, Korchin,
  and Koch}}]{scherer}
\bibinfo{author}{\bibfnamefont{S.}~\bibnamefont{Scherer}},
  \bibinfo{author}{\bibfnamefont{A.~Y.} \bibnamefont{Korchin}},
  \bibnamefont{and} \bibinfo{author}{\bibfnamefont{J.~H.} \bibnamefont{Koch}},
  \bibinfo{journal}{Phys. Rev. C} \textbf{\bibinfo{volume}{54}},
  \bibinfo{pages}{904} (\bibinfo{year}{1996}).

\bibitem[{\citenamefont{Metz and Drechsel}(1996)}]{metz}
\bibinfo{author}{\bibfnamefont{A.}~\bibnamefont{Metz}} \bibnamefont{and}
  \bibinfo{author}{\bibfnamefont{D.}~\bibnamefont{Drechsel}},
  \bibinfo{journal}{Z. Phys. A} \textbf{\bibinfo{volume}{356}},
  \bibinfo{pages}{351} (\bibinfo{year}{1996}).

\bibitem[{\citenamefont{Pasquini and Boffi}(1996)}]{pasquini}
\bibinfo{author}{\bibfnamefont{B.}~\bibnamefont{Pasquini}} \bibnamefont{and}
  \bibinfo{author}{\bibfnamefont{S.}~\bibnamefont{Boffi}},
  \bibinfo{journal}{Phys. Lett. B} \textbf{\bibinfo{volume}{386}},
  \bibinfo{pages}{29} (\bibinfo{year}{1996}).

\bibitem[{\citenamefont{Drechsel et~al.}(1997)\citenamefont{Drechsel,
  Knochlein, Metz, and Scherer}}]{drechsel}
\bibinfo{author}{\bibfnamefont{D.}~\bibnamefont{Drechsel}},
  \bibinfo{author}{\bibfnamefont{G.}~\bibnamefont{Knochlein}},
  \bibinfo{author}{\bibfnamefont{A.}~\bibnamefont{Metz}}, \bibnamefont{and}
  \bibinfo{author}{\bibfnamefont{S.}~\bibnamefont{Scherer}},
  \bibinfo{journal}{Phys. Rev. C} \textbf{\bibinfo{volume}{55}},
  \bibinfo{pages}{424} (\bibinfo{year}{1997}).

\bibitem[{\citenamefont{Drechsel et~al.}(1998)\citenamefont{Drechsel,
  Knochlein, Korchin, Metz, and Scherer}}]{drechsel2}
\bibinfo{author}{\bibfnamefont{D.}~\bibnamefont{Drechsel}},
  \bibinfo{author}{\bibfnamefont{G.}~\bibnamefont{Knochlein}},
  \bibinfo{author}{\bibfnamefont{A.~Y.} \bibnamefont{Korchin}},
  \bibinfo{author}{\bibfnamefont{A.}~\bibnamefont{Metz}}, \bibnamefont{and}
  \bibinfo{author}{\bibfnamefont{S.}~\bibnamefont{Scherer}},
  \bibinfo{journal}{Phys. Rev. C} \textbf{\bibinfo{volume}{57}},
  \bibinfo{pages}{941} (\bibinfo{year}{1998}).

\bibitem[{\citenamefont{Hemmert et~al.}(1997)\citenamefont{Hemmert, Holstein,
  Knochlein, and Scherer}}]{HHKS}
\bibinfo{author}{\bibfnamefont{T.~R.} \bibnamefont{Hemmert}},
  \bibinfo{author}{\bibfnamefont{B.~R.} \bibnamefont{Holstein}},
  \bibinfo{author}{\bibfnamefont{G.}~\bibnamefont{Knochlein}},
  \bibnamefont{and} \bibinfo{author}{\bibfnamefont{S.}~\bibnamefont{Scherer}},
  \bibinfo{journal}{Phys. Rev. Lett.} \textbf{\bibinfo{volume}{79}},
  \bibinfo{pages}{22} (\bibinfo{year}{1997}).

\bibitem[{\citenamefont{Hemmert et~al.}(2000)\citenamefont{Hemmert, Holstein,
  Knochlein, and Drechsel}}]{HHKD}
\bibinfo{author}{\bibfnamefont{T.~R.} \bibnamefont{Hemmert}},
  \bibinfo{author}{\bibfnamefont{B.~R.} \bibnamefont{Holstein}},
  \bibinfo{author}{\bibfnamefont{G.}~\bibnamefont{Knochlein}},
  \bibnamefont{and} \bibinfo{author}{\bibfnamefont{D.}~\bibnamefont{Drechsel}},
  \bibinfo{journal}{Phys. Rev. D} \textbf{\bibinfo{volume}{62}},
  \bibinfo{pages}{014013} (\bibinfo{year}{2000}).

\bibitem[{\citenamefont{L'vov et~al.}(2001)\citenamefont{L'vov, Scherer,
  Pasquini, Unkmeir, and Drechsel}}]{lvov}
\bibinfo{author}{\bibfnamefont{A.~I.} \bibnamefont{L'vov}},
  \bibinfo{author}{\bibfnamefont{S.}~\bibnamefont{Scherer}},
  \bibinfo{author}{\bibfnamefont{B.}~\bibnamefont{Pasquini}},
  \bibinfo{author}{\bibfnamefont{C.}~\bibnamefont{Unkmeir}}, \bibnamefont{and}
  \bibinfo{author}{\bibfnamefont{D.}~\bibnamefont{Drechsel}},
  \bibinfo{journal}{Phys. Rev. C} \textbf{\bibinfo{volume}{64}},
  \bibinfo{pages}{015203} (\bibinfo{year}{2001}).

\bibitem[{\citenamefont{Gorchtein}(2010)}]{gorchtein_vcs}
\bibinfo{author}{\bibfnamefont{M.}~\bibnamefont{Gorchtein}},
  \bibinfo{journal}{Phys. Rev. C} \textbf{\bibinfo{volume}{81}},
  \bibinfo{pages}{015206} (\bibinfo{year}{2010}).

\bibitem[{\citenamefont{Ho and Scherrer}(2013)}]{anapole_dm1}
\bibinfo{author}{\bibfnamefont{C.~M.} \bibnamefont{Ho}} \bibnamefont{and}
  \bibinfo{author}{\bibfnamefont{R.~J.} \bibnamefont{Scherrer}},
  \bibinfo{journal}{Phys. Lett.} \textbf{\bibinfo{volume}{B722}},
  \bibinfo{pages}{341} (\bibinfo{year}{2013}).

\bibitem[{\citenamefont{Gao et~al.}(2014)\citenamefont{Gao, Ho, and
  Scherrer}}]{anapole_dm2}
\bibinfo{author}{\bibfnamefont{Y.}~\bibnamefont{Gao}},
  \bibinfo{author}{\bibfnamefont{C.~M.} \bibnamefont{Ho}}, \bibnamefont{and}
  \bibinfo{author}{\bibfnamefont{R.~J.} \bibnamefont{Scherrer}},
  \bibinfo{journal}{Phys. Rev.} \textbf{\bibinfo{volume}{D89}},
  \bibinfo{pages}{045006} (\bibinfo{year}{2014}).

\bibitem[{\citenamefont{Anchordoqui et~al.}(2015)\citenamefont{Anchordoqui,
  Barger, Goldberg, Huang, Marfatia, da~Silva, and Weiler}}]{Anchordoqui}
\bibinfo{author}{\bibfnamefont{L.~A.} \bibnamefont{Anchordoqui}},
  \bibinfo{author}{\bibfnamefont{V.}~\bibnamefont{Barger}},
  \bibinfo{author}{\bibfnamefont{H.}~\bibnamefont{Goldberg}},
  \bibinfo{author}{\bibfnamefont{X.}~\bibnamefont{Huang}},
  \bibinfo{author}{\bibfnamefont{D.}~\bibnamefont{Marfatia}},
  \bibinfo{author}{\bibfnamefont{L.~H.~M.} \bibnamefont{da~Silva}},
  \bibnamefont{and} \bibinfo{author}{\bibfnamefont{T.~J.}
  \bibnamefont{Weiler}}, \bibinfo{journal}{Phys. Rev. D}
  \textbf{\bibinfo{volume}{92}}, \bibinfo{pages}{063504}
  (\bibinfo{year}{2015}).

\bibitem[{\citenamefont{Chua and Wong}(2016)}]{chua}
\bibinfo{author}{\bibfnamefont{C.-K.} \bibnamefont{Chua}} \bibnamefont{and}
  \bibinfo{author}{\bibfnamefont{G.-G.} \bibnamefont{Wong}},
  \bibinfo{journal}{Physical Review D} \textbf{\bibinfo{volume}{94}},
  \bibinfo{pages}{035002} (\bibinfo{year}{2016}).

\bibitem[{\citenamefont{Kumar et~al.}(2016)\citenamefont{Kumar, Sandick, Teng,
  and Yamamoto}}]{kumar}
\bibinfo{author}{\bibfnamefont{J.}~\bibnamefont{Kumar}},
  \bibinfo{author}{\bibfnamefont{P.}~\bibnamefont{Sandick}},
  \bibinfo{author}{\bibfnamefont{F.}~\bibnamefont{Teng}}, \bibnamefont{and}
  \bibinfo{author}{\bibfnamefont{T.}~\bibnamefont{Yamamoto}},
  \bibinfo{journal}{Phys. Rev. D} \textbf{\bibinfo{volume}{94}},
  \bibinfo{pages}{015022} (\bibinfo{year}{2016}).

\bibitem[{\citenamefont{Ahmed et~al.}(2018)\citenamefont{Ahmed, Duch,
  Grzadkowski, and Iglicki}}]{ahmed}
\bibinfo{author}{\bibfnamefont{A.}~\bibnamefont{Ahmed}},
  \bibinfo{author}{\bibfnamefont{M.}~\bibnamefont{Duch}},
  \bibinfo{author}{\bibfnamefont{B.}~\bibnamefont{Grzadkowski}},
  \bibnamefont{and} \bibinfo{author}{\bibfnamefont{M.}~\bibnamefont{Iglicki}},
  \bibinfo{journal}{Eur. Phys. J. C} \textbf{\bibinfo{volume}{78}},
  \bibinfo{pages}{905} (\bibinfo{year}{2018}).

\bibitem[{\citenamefont{Neves et~al.}(2021)\citenamefont{Neves, Okada, and
  Okada}}]{neves2021majorana}
\bibinfo{author}{\bibfnamefont{M.~J.} \bibnamefont{Neves}},
  \bibinfo{author}{\bibfnamefont{N.}~\bibnamefont{Okada}}, \bibnamefont{and}
  \bibinfo{author}{\bibfnamefont{S.}~\bibnamefont{Okada}},
  \bibinfo{journal}{arXiv:2103.08873}  (\bibinfo{year}{2021}).

\bibitem[{\citenamefont{Latimer}(2017)}]{anapole_2photon}
\bibinfo{author}{\bibfnamefont{D.~C.} \bibnamefont{Latimer}},
  \bibinfo{journal}{Phys. Rev. D} \textbf{\bibinfo{volume}{95}},
  \bibinfo{pages}{095023} (\bibinfo{year}{2017}).

\bibitem[{\citenamefont{Denner et~al.}(1992{\natexlab{a}})\citenamefont{Denner,
  Eck, Hahn, and Kublbeck}}]{denner1}
\bibinfo{author}{\bibfnamefont{A.}~\bibnamefont{Denner}},
  \bibinfo{author}{\bibfnamefont{H.}~\bibnamefont{Eck}},
  \bibinfo{author}{\bibfnamefont{O.}~\bibnamefont{Hahn}}, \bibnamefont{and}
  \bibinfo{author}{\bibfnamefont{J.}~\bibnamefont{Kublbeck}},
  \bibinfo{journal}{Phys. Lett.} \textbf{\bibinfo{volume}{B291}},
  \bibinfo{pages}{278} (\bibinfo{year}{1992}{\natexlab{a}}).

\bibitem[{\citenamefont{Denner et~al.}(1992{\natexlab{b}})\citenamefont{Denner,
  Eck, Hahn, and Kublbeck}}]{denner2}
\bibinfo{author}{\bibfnamefont{A.}~\bibnamefont{Denner}},
  \bibinfo{author}{\bibfnamefont{H.}~\bibnamefont{Eck}},
  \bibinfo{author}{\bibfnamefont{O.}~\bibnamefont{Hahn}}, \bibnamefont{and}
  \bibinfo{author}{\bibfnamefont{J.}~\bibnamefont{Kublbeck}},
  \bibinfo{journal}{Nucl. Phys.} \textbf{\bibinfo{volume}{B387}},
  \bibinfo{pages}{467} (\bibinfo{year}{1992}{\natexlab{b}}).

\bibitem[{\citenamefont{Patel}(2017)}]{patel}
\bibinfo{author}{\bibfnamefont{H.~H.} \bibnamefont{Patel}},
  \bibinfo{journal}{Comput.~Phys.~Commun.} \textbf{\bibinfo{volume}{218}},
  \bibinfo{pages}{66} (\bibinfo{year}{2017}).

\bibitem[{\citenamefont{Bernab\'eu et~al.}(2000)\citenamefont{Bernab\'eu,
  Cabral-Rosetti, Papavassiliou, and Vidal}}]{neutrino_anapole}
\bibinfo{author}{\bibfnamefont{J.}~\bibnamefont{Bernab\'eu}},
  \bibinfo{author}{\bibfnamefont{L.~G.} \bibnamefont{Cabral-Rosetti}},
  \bibinfo{author}{\bibfnamefont{J.}~\bibnamefont{Papavassiliou}},
  \bibnamefont{and} \bibinfo{author}{\bibfnamefont{J.}~\bibnamefont{Vidal}},
  \bibinfo{journal}{Phys. Rev. D} \textbf{\bibinfo{volume}{62}},
  \bibinfo{pages}{113012} (\bibinfo{year}{2000}).

\bibitem[{\citenamefont{Kolb and Turner}(1990)}]{kolb_turner}
\bibinfo{author}{\bibfnamefont{E.}~\bibnamefont{Kolb}} \bibnamefont{and}
  \bibinfo{author}{\bibfnamefont{M.}~\bibnamefont{Turner}},
  \emph{\bibinfo{title}{The Early Universe}}
  (\bibinfo{publisher}{Addison-Wesley}, \bibinfo{address}{Redwood City, CA},
  \bibinfo{year}{1990}).

\bibitem[{\citenamefont{Scherrer and Turner}(1986)}]{Scherrer:1985zt}
\bibinfo{author}{\bibfnamefont{R.~J.} \bibnamefont{Scherrer}} \bibnamefont{and}
  \bibinfo{author}{\bibfnamefont{M.~S.} \bibnamefont{Turner}},
  \bibinfo{journal}{Phys. Rev.} \textbf{\bibinfo{volume}{D33}},
  \bibinfo{pages}{1585} (\bibinfo{year}{1986}), \bibinfo{note}{[Erratum: Phys.
  Rev. {\bf D34}, 3263 (1986)]}.

\bibitem[{\citenamefont{{P.~A.~Zyla, {\em et al.} (Particle Data
  Group)}}(2020)}]{pdg2020}
\bibinfo{author}{\bibnamefont{{P.~A.~Zyla, {\em et al.} (Particle Data
  Group)}}}, \bibinfo{journal}{Prog.~Theor.~Exp.~Phys.~}
  \textbf{\bibinfo{volume}{2020}}, \bibinfo{pages}{083C01}
  (\bibinfo{year}{2020}).

\bibitem[{\citenamefont{Alves et~al.}(2018)\citenamefont{Alves, Santos, and
  Sinha}}]{alves}
\bibinfo{author}{\bibfnamefont{A.}~\bibnamefont{Alves}},
  \bibinfo{author}{\bibfnamefont{A.~C.~O.} \bibnamefont{Santos}},
  \bibnamefont{and} \bibinfo{author}{\bibfnamefont{K.}~\bibnamefont{Sinha}},
  \bibinfo{journal}{Phys. Rev. D} \textbf{\bibinfo{volume}{97}},
  \bibinfo{pages}{055023} (\bibinfo{year}{2018}).

\bibitem[{\citenamefont{Turner and Widrow}(1988)}]{PhysRevD.37.2743}
\bibinfo{author}{\bibfnamefont{M.~S.} \bibnamefont{Turner}} \bibnamefont{and}
  \bibinfo{author}{\bibfnamefont{L.~M.} \bibnamefont{Widrow}},
  \bibinfo{journal}{Phys. Rev. D} \textbf{\bibinfo{volume}{37}},
  \bibinfo{pages}{2743} (\bibinfo{year}{1988}).

\bibitem[{\citenamefont{Sigl et~al.}(1997)\citenamefont{Sigl, Olinto, and
  Jedamzik}}]{PhysRevD.55.4582}
\bibinfo{author}{\bibfnamefont{G.}~\bibnamefont{Sigl}},
  \bibinfo{author}{\bibfnamefont{A.~V.} \bibnamefont{Olinto}},
  \bibnamefont{and} \bibinfo{author}{\bibfnamefont{K.}~\bibnamefont{Jedamzik}},
  \bibinfo{journal}{Phys. Rev. D} \textbf{\bibinfo{volume}{55}},
  \bibinfo{pages}{4582} (\bibinfo{year}{1997}).

\bibitem[{\citenamefont{Baym et~al.}(1996)\citenamefont{Baym, B\"odeker, and
  McLerran}}]{PhysRevD.53.662}
\bibinfo{author}{\bibfnamefont{G.}~\bibnamefont{Baym}},
  \bibinfo{author}{\bibfnamefont{D.}~\bibnamefont{B\"odeker}},
  \bibnamefont{and} \bibinfo{author}{\bibfnamefont{L.}~\bibnamefont{McLerran}},
  \bibinfo{journal}{Phys. Rev. D} \textbf{\bibinfo{volume}{53}},
  \bibinfo{pages}{662} (\bibinfo{year}{1996}).

\bibitem[{\citenamefont{{Quashnock} et~al.}(1989)\citenamefont{{Quashnock},
  {Loeb}, and {Spergel}}}]{QCDtrans_currents}
\bibinfo{author}{\bibfnamefont{J.~M.} \bibnamefont{{Quashnock}}},
  \bibinfo{author}{\bibfnamefont{A.}~\bibnamefont{{Loeb}}}, \bibnamefont{and}
  \bibinfo{author}{\bibfnamefont{D.~N.} \bibnamefont{{Spergel}}},
  \bibinfo{journal}{Astrophys.~J.~Lett.~} \textbf{\bibinfo{volume}{344}},
  \bibinfo{pages}{L49} (\bibinfo{year}{1989}).

\bibitem[{\citenamefont{Cheng and Olinto}(1994)}]{PhysRevD.50.2421}
\bibinfo{author}{\bibfnamefont{B.}~\bibnamefont{Cheng}} \bibnamefont{and}
  \bibinfo{author}{\bibfnamefont{A.~V.} \bibnamefont{Olinto}},
  \bibinfo{journal}{Phys. Rev. D} \textbf{\bibinfo{volume}{50}},
  \bibinfo{pages}{2421} (\bibinfo{year}{1994}).

\bibitem[{\citenamefont{de~Souza and Opher}(2008)}]{PhysRevD.77.043529}
\bibinfo{author}{\bibfnamefont{R.~S.} \bibnamefont{de~Souza}} \bibnamefont{and}
  \bibinfo{author}{\bibfnamefont{R.}~\bibnamefont{Opher}},
  \bibinfo{journal}{Phys. Rev. D} \textbf{\bibinfo{volume}{77}},
  \bibinfo{pages}{043529} (\bibinfo{year}{2008}).

\bibitem[{\citenamefont{Siegel and Fry}(2006)}]{siegel_fry}
\bibinfo{author}{\bibfnamefont{E.~R.} \bibnamefont{Siegel}} \bibnamefont{and}
  \bibinfo{author}{\bibfnamefont{J.~N.} \bibnamefont{Fry}}
  (\bibinfo{year}{2006}),
  \urlprefix\url{https://arxiv.org/abs/astro-ph/0609031}.

\bibitem[{\citenamefont{Purcell}(1946)}]{purcell_spontaneous_1946}
\bibinfo{author}{\bibfnamefont{E.~M.} \bibnamefont{Purcell}},
  \bibinfo{journal}{Physical Review} \textbf{\bibinfo{volume}{69}},
  \bibinfo{pages}{681} (\bibinfo{year}{1946}).

\end{thebibliography}

\end{document}